\documentclass{iucr}

\usepackage{siunitx}
\usepackage{color}
\usepackage{mathtools}
\usepackage[normalem]{ulem}
\usepackage{adjustbox}
\usepackage{array}
\usepackage{booktabs}
\usepackage{multirow}

\usepackage{amsmath}
\usepackage[hyphens,spaces,obeyspaces]{url}

\DeclareMathOperator{\sinc}{sinc}

\definecolor{JSR_blue}{RGB}{51, 102, 154}
\newcommand{\jsrblue}[1]{\textcolor{JSR_blue}{#1}}

\newcolumntype{R}[2]{%
    >{\adjustbox{angle=#1,lap=\width-(#2)}\bgroup}%
    l%
    <{\egroup}%
}

\makeatletter
\@ifclasswith{iucr}{preprint}{

}{

}
\makeatother

 \journalcode{X}              

\begin{document}                  


\title{Modelling undulators in ray tracing simulations}

\cauthor[]{\jsrblue{Manuel}}{\jsrblue{Sanchez del Rio}}{srio@esrf.eu}{address if different from \aff}
\author[]{\jsrblue{Juan}}{\jsrblue{Reyes-Herrera}}

\aff[]{ESRF - The European Synchrotron, 71 Avenue des Martyrs, 38000 Grenoble, \country{France}}

\maketitle                        


\begin{synopsis}
We describe the theory and models implemented in the undulator sources of the ray tracing code SHADOW4.  
\end{synopsis}


\begin{abstract}
We introduce a model that can accurately simulate radiation from undulator sources for ray tracing applications. It incorporates several key effects relevant to 4$^\text{th}$ generation synchrotron sources, such as electron emittance, energy spread, and diffraction-limited beam size. This code has been developed as part of SHADOW4, the latest version of the widely used SHADOW X-ray optics ray tracing program. 
The approach relies on calculating the field distribution in the far field, which determines the ray divergences. The integration of existing models for electron energy spread is also addressed. Rays sampled at the source follow a size distribution derived by backpropagating the far field radiation. These models are detailed, and several examples are provided.

\end{abstract}

\section{Introduction}
\label{sec:introduction}

Undulators are the most popular magnetic structures for producing synchrotron radiation in third and fourth-generation sources.
The radiated beam by an undulator is usually more brilliant than the other sources: it is much more collimated than in wigglers and bending magnets, and it is as intense or more than the wiggler at certain photon energies.
The theory of the undulator radiation (UR) is well understood and several comprehensive texts are available \cite{BookDuke, elleaume, BookClarke}.
The radiation emitted by the undulator  exhibits distinct structures both in its spectrum, presenting peaks at some photon energies (resonances), and in its geometry (wavefront size and derived divergences). 
Sseveral software tools are available to compute the characteristics of the UR. Among them, SRW \cite{codeSRW} and SPECTRA \cite{Tanaka2001}  are the most advanced. 

Ray tracing packages create undulator sources by sampling rays according to the distributions given by the undulator theory. SHADOW, since its first version SHADOW1 \cite{Cerrina1984}, included an undulator model \cite{shadow2undulators}.
In SHADOW3 \cite{codeSHADOW} and its ShadowOui interface \cite{codeSHADOWOUI}, the undulator calculations were refactored, and partially replaced by new python code. Moreover, ShadowOui also provides an ``Undulator Gaussian" application, that creates a source with rays that follow Gaussian distribution that approximate the undulator distributions.
This has been found very useful when in a first phase or prototyping beamlines using undulators as sources. 
In SHADOW4 \cite{ShadowSRN2023}, the newly refactored and enhanced version of this popular ray tracing code, we have reimplemented, improved, corrected, and upgraded the undulator algorithms, significantly improving the performance and accuracy of ray tracing simulations.
Special attention has been given to accurately implement certain features that have become crucial with the emergence of 4$^\text{th}$ generation synchrotron sources. 
Notably, this includes the impact of electron energy spread, which is particularly important when utilizing radiation at high harmonics, as well as the accurate description of diffraction-limited beam size.

This paper aims to describe the methods and algorithms used in SHADOW4 for simulating undulator sources.
In a first section we summarize the most important results of the undulator theory used, with a detailed discussion of the Gaussian approximations for beam sizes and divergences.
Next, the algorithms and methods for sampling rays from undulator sources, both in the Gaussian approximation and in the full model, are described. Finally, several examples are provided, followed by a discussion. 

\section{Summary of the theory of undulator emission}
\label{sec:undulator}

The undulator magnets induce in the electrons a periodic (mostly sinusoidal) trajectory.
The small deflection of each electron at each oscillation of the magnetic field makes it possible that the photons produced in a crest of the electron periodic trajectory interfere with the photons originated from the next oscillation crest, thus producing radiation with non-smooth characteristics (spectral and spatial).
The spectrum contains peaks at photon energies proportional to the so-called resonance.
It depends on the deflection parameter $K$.
The $K$ value for an electron traveling in an oscillating magnetic field $B \cos(2 \pi z/\lambda_u)$ (with $z$ the spatial coordinate along the undulator, $B$ the maximum magnetic field, and $\lambda_u$ the undulator period) is

\begin{equation}
K = \frac{e B \lambda_u}{2 \pi m c} \sim 93.3729 B[T] \lambda_u[m],
\label{eq:K}
\end{equation}
with $m$ and $e$ the mass and charge, respectively, of the electron, $c$ the velocity of the light. The resonance is found at the photon wavelength
\begin{equation}
\lambda = \frac{1+K^2/2+\gamma^2\theta^2}{2 \gamma^2} \lambda_u,
\label{eq:resonance}
\end{equation}
with 
$\theta$ the observation angle ($\theta$=0 on-axis) and $\gamma$$\approx$1957$\mathcal{E}$$[\mathrm{GeV}]$, the Lorentz factor with $\mathcal{E}$ the electron energy.

\subsection{Emission from a relativistic electron}

An ultrarelativistic charged particle traveling along a curved, often wiggly trajectory, typically generated by alternating magnetic fields in insertion devices (ID), emits radiation.
The electric field can be calculated in the framework of classical electrodynamics [see e.g. equation~(14.14) in \cite{jackson}]. 
The electric field at an observation point $\textbf{r}=(x,y,z)$ is proportional to this integral along the electron trajectory
\begin{equation}
\begin{split}
    &\int_{-\infty}^{\infty}
    \biggl[ 
    \frac{\textbf{n} \times [(\textbf{n} - \mathbf{\beta}) \times \dot{\mathbf{\beta}}]}
    {(1- \mathbf{\beta} \cdot \textbf{n})^3} +\\
    &\qquad+\frac{c}{\gamma^2 R}   \frac{\textbf{n} - \mathbf{\beta}}{(1-\mathbf{\beta} \cdot \textbf{n})^3} \biggr]
    \exp[i \omega (t - \textbf{n}\cdot\textbf{r}/c)] \mathrm{d}t
\end{split}\label{eq:undulator}
\end{equation}
where $c$ the velocity of light, $\omega$ is the radiated frequency, $\mathbf{\beta}\text{=}\dot{\mathbf{r}}\big/c$ is the electron relative velocity, 
and the dot denotes the time derivative.
Also $\textbf{n}(t)\text{=}[\textbf{r}-\textbf{r}_{\textbf{e}}(t)]\big/|\textbf{r}-\textbf{r}_{\textbf{e}}(t)|$ is the unit vector pointing from the particle to the observation point $\textbf{r}$; the electron trajectory is represented by $\textbf{r}_{\textbf{e}}(t)$, which is completely determined by the 3D distribution of the magnetic field inside the ID and the electron initial conditions prior to entering it.
The origin of the vector $\textbf{r}$ is usually at the centre of the ID/straight section.

Equation~(\ref{eq:undulator}) describes a fully spatially-coherent field emitted by a single electron. 
In an idealized zero-emittance storage ring, the electrons follow a ``filament beam". It contains $N_e$ electrons that follow exactly the same trajectory $\textbf{r}_{\textbf{e}}(t)$, therefore the radiation intensity, calculated from the square of equation~(\ref{eq:undulator}), will be affected by a multiplicative factor $N_e$, or for practical effects will be expressed as a function of the electron current. 

Several codes are available in the synchrotron community to calculate the undulator emission characteristics in different cases.
The codes URGENT \cite{codeURGENT} and US \cite{codeUS} compute undulator emission in the far-field for undulators with a sinusoidal magnetic field. 
The codes SRW \cite{codeSRW} and SPECTRA \cite{Tanaka2001} are more generic as they calculate emission in the near and far-field for any electron trajectory (with different initial conditions) and submitted to an arbitrary magnetic field.
We use pySRU \cite{pySRU}, an open source code developed in Python, that calculates the emission using equation~(\ref{eq:undulator}).
It is well integrated in python ecosystems, such as OASYS \cite{codeOASYS}, which includes the SHADOW4 user interface.
Portions of pySRU have been incorporated into the internal code of SHADOW4.  

The flux spectrum $F(E)$, with $F$ the flux in photons/s/0.1\%bw and $E\text{ = }\hbar\omega$ the photon energy (in eV) is obtained by fixing a coordinate $z$ (the distance from the center of the undulator to an observation plane) and integrating over the $x,y$ variables ($x$ is in the horizontal plane and $y$ in the vertical plane\footnote{In SHADOW4 the axes naming is different: the propagation is along $y$, and at the source level $x$ is horizontal and $z$ vertical.}).
The spectrum ``on-axis" [i.e. integrated over an infinitesimal interval of $(x,y)$] presents peaks at energies corresponding to the values in equation~(\ref{eq:resonance}) (with $\theta$=0).
These peaks have the form of a $\sinc^2(x)$ function with $x\text{=}\pi N (\frac{E}{E_0}-n)$ \cite{elleaumeChapter3} ($N$ is the number of undulator periods, $E_0$ is the resonance energy, and $n$ is the harmonic number).
As far as one opens the integration range in $\theta$ or in $(x,y)$ (or in other words, we open an acceptance slit) the peaks become larger because the resonance shifts with the conservation angle $\theta$ in equation~(\ref{eq:resonance}).

The intensity map of the radiation at a $z$ sufficiently large (far field) does not change its shape but it only expands with $z$.
We can then speak about ``divergence'', in terms of the radial angle $\theta\text{=}(x^2+y^2)^{1/2}/z$ (the horizontal angle is $\theta_x\text{=}x/z$ and the vertical one $\theta_y\text{=}y/z$). Under some approximations (far field, sinusoidal field, on-resonance) the intensity reduces to [equations~23-24 in \cite{elleaumeChapter3}]
\begin{equation}\label{eq:elleaumetheorydivergence}
    I(\bar{\theta}) \propto 
    \sinc^2( \frac{\pi}{2} \bar{\theta}^2 ),
\end{equation}
with $\bar{\theta}\text{ = }\theta (L/\lambda)^{1/2}$, $L$ is the undulator length and $\lambda$ the photon wavelength at a given harmonic.

To obtain the source size (intensity map at $z$ = 0), the equation~(\ref{eq:undulator}) can not be directly applied to points within the electron trajectory. Therefore, it is necessary to compute the electric field in a $z$ point external to the undulator and then backpropagate it (using Fresnel or Fraunhofer propagators) to the plane at $z$ = 0. Using as before some approximations the source size can be expressed as a Hankel transform  (the Fraunhofer propagator in radial coordinates) of equation~(\ref{eq:elleaumetheorydivergence}) which gives [equation~29 in \cite{elleaumeChapter3}]:
\begin{equation}\label{eq:elleaumetheorysize}
I(\bar{r}) \propto \int{\phi J_0(\phi \bar r ) d\phi},
\end{equation}
where $J_0(x)$ is the Bessel function of the first kind and zeroth-order, and $\bar r$ is the ``reduced" radial coordinate $\bar r \text{ = } 2 \pi r (2 \lambda L)^{-1/2}$.
\subsection{Gaussian approximation of undulator size and divergence at resonance}
\label{sec:electronbeam}

The divergence or angular  distribution of the UR can be calculated by representing the flux (F) as a function of the horizontal ($\theta_x$) and vertical ($\theta_y$) angles, or the radial angle ($\theta\text{ =}(\theta_x+\theta_y)^{1/2}$). Near resonance (and its odd harmonics), the distribution displays a pronounced peak, known as the "central cone," along with some surrounding rings.
At exact resonance, the distribution is described by equation~(\ref{eq:elleaumetheorydivergence}).


As far as the photon energy is reduced (red-shifted), the peak opens a hole in the middle that separates into two or more peaks to eventually disappear. 
The width of the intensity profile of this radiation cone is a fundamental parameter for researchers and engineers working with synchrotrons. 
Different works found in literature use different approximations under multiple hypotheses with some discrepancy in the results.
We summarize them from a historic perspective and identify the equations implemented in SHADOW4.

\begin{figure}
   \centering
   \includegraphics[width=0.49\textwidth]{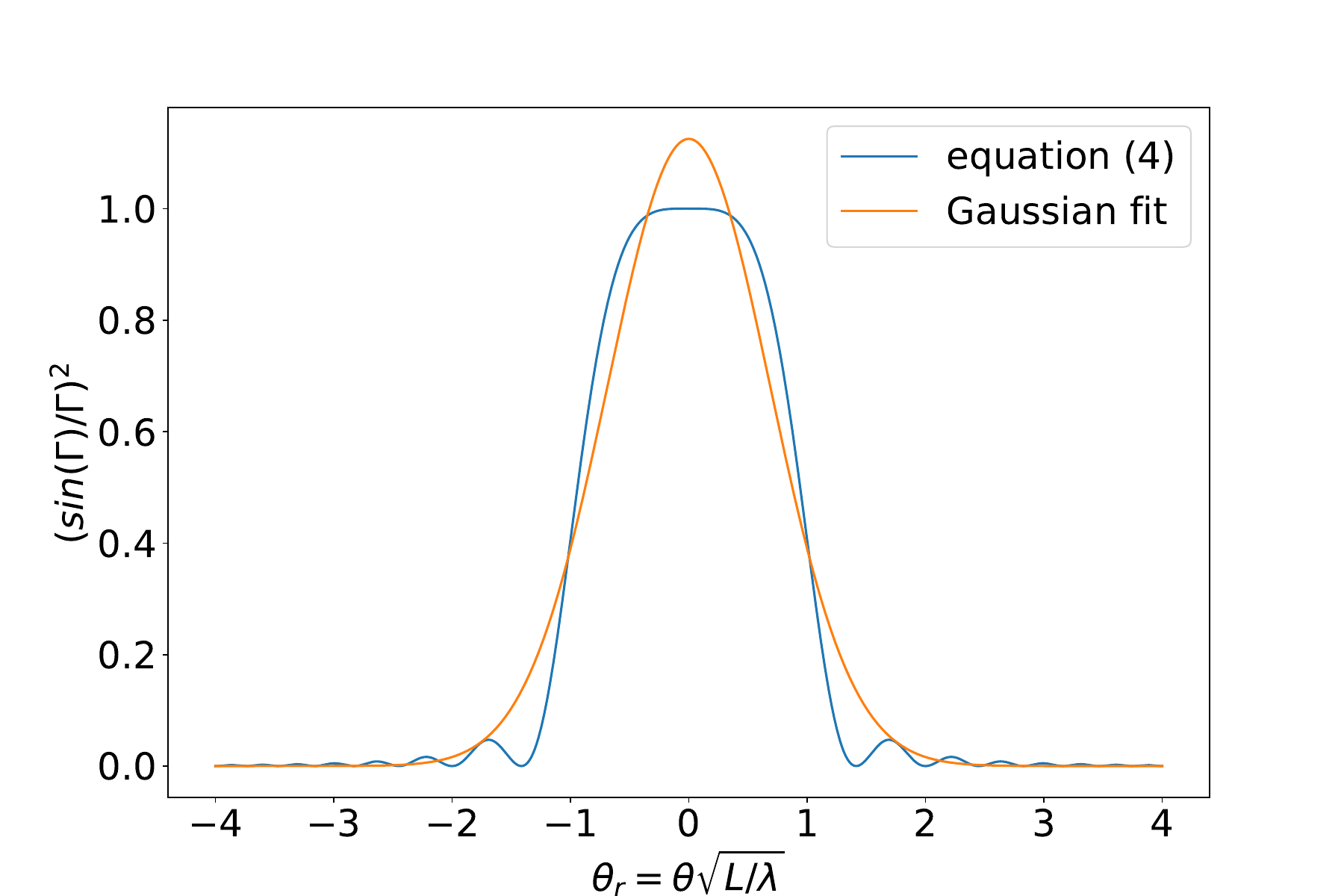}
   \includegraphics[width=0.49\textwidth]{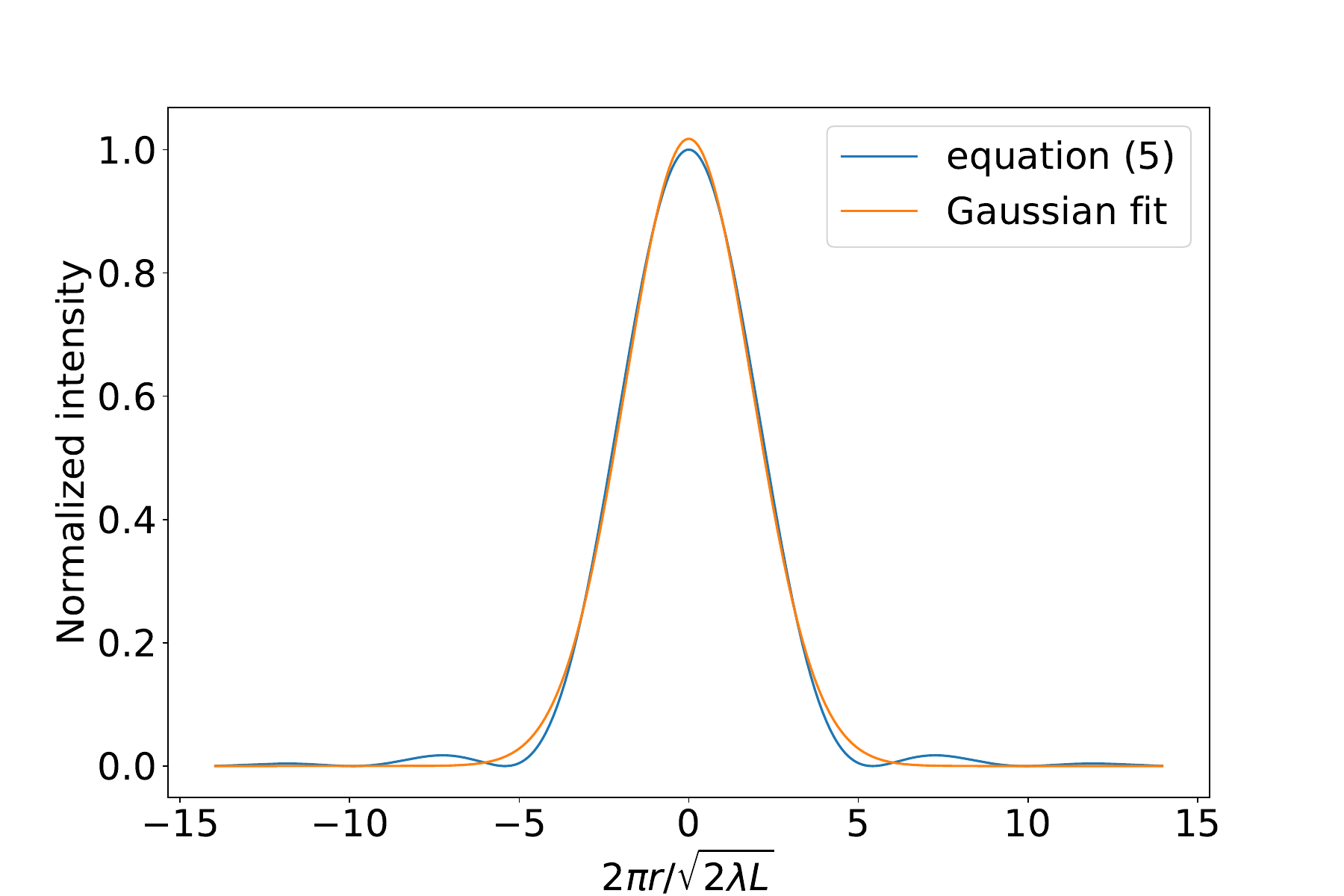}
   
   \caption{Gaussian fits of the intensity versus reduced emission angle (a) and reduced size (b) at the center of the undulator plane,  after Ref.~\cite{elleaume}.}
   \label{fig:undulator_fit}
\end{figure}

The natural divergence of synchrotron light for all sources (bending magnets, wigglers and undulators) is approximately proportional to $\gamma^{-1}$. For undulator sources, it is smaller by a factor that depends on the number of undulator periods. Using simple arguments ~\cite{krinsky}  affirms that the angular broadening of the radiation is defined as: 
\begin{equation}
\sigma_{r'} \cong \frac{1}{\gamma}\sqrt{ \frac{(1+K^2/2)}{2Nn}}=\sqrt{\frac{\lambda}{L} }
\label{eq:kim_sigmaprime}
\end{equation}
The same expression is used by Kim in \cite{kim1986a,kim1986b} and also in the X-ray Data Booklet \cite{xraydatabooklet} or in \cite{BookDuke} [Eq.~(14.21)]. Notice that in the original texts the width is said to be ``half width", which in principle is different from a ``sigma" $\sigma$ in a Gaussian distribution, which has a full width at half maximum (FWHM) FWHM = $2 \sqrt{2 \ln 2} \sigma \sim 2.355 \sigma$. In Kim's papers from 1989 \cite{kim1989}, the Gaussian approximation is obtained by matching its integral with the angular distribution of intensity (the $\sinc$ function in equation~\ref{eq:elleaumetheorydivergence}) and obtaining a smaller divergence:
\begin{equation}
\sigma_{r'} = \frac{1}{2\gamma}\sqrt{ \frac{(1+K^2/2)}{Nn}}=\sqrt{\frac{\lambda}{2L} }
\label{eq:kimnew_sigmaprime}
\end{equation}
Elleaume \cite{elleaumeChapter3} performs a Gaussian fit on the intensity versus emission angle at the resonance [equation~(\ref{eq:elleaumetheorydivergence})] and obtained: 
\begin{equation}
\sigma_{r'} = 0.69 \sqrt{\frac{\lambda}{L}}.
\label{eq:elleaume_sigmaprime}
\end{equation}
Note that in this equation the numeric factor 0.69 is very close  to $1/\sqrt{2}=0.707$ in equation~(\ref{eq:kimnew_sigmaprime}). Therefore, if not identical, they are in close agreement (within 2\%). We repeated this fit with the same result [see Fig.~(\ref{fig:undulator_fit}a)].
However, one can remark by a simple visual inspection that the fit is not good: the intensity profile is far from being Gaussian.
Another practical method to obtain the $\sigma$ value is to compute the root mean square (r.m.s.) of the intensity distribution, identical to the standard deviation because the mean is zero. This is practical for numeric calculations, but may end in infinite values using the theoretical equations, as discussed in \cite{elleaumeChapter3}. 
Therefore, care must be taken when using these approximated values. 
In recent bibliography, most authors agree with equation~(\ref{eq:kimnew_sigmaprime}), like \cite{Tanaka2014, Walker2019}.


The situation is more controversial when discussing the radiation size $\sigma_r$, or the spatial with of the radiation at the center of the undulator. Before summarizing the bibliographic results, we remind that a Gaussian beam (the first mode of a Gaussian Shell-model beam) verifies
\begin{equation}
\sigma_r \sigma_{r'} = \frac{\lambda}{4 \pi}.
\label{eq:sigmasigmaprime-gaussianbeam}
\end{equation}

Kim [(Eq.~21 in \cite{kim1986b} or Eq.~6.37 in \cite{kim1989})] supposed that the UR verifies equation~(\ref{eq:sigmasigmaprime-gaussianbeam}). Therefore, he implicitly assumed the validity of approximating UR by a Gaussian Shell-model, or, in other words, accepting that the emission at the resonance is Gaussian. In consequence, depending on the divergence value used two results are found: Combining equations (\ref{eq:sigmasigmaprime-gaussianbeam}) and (\ref{eq:kim_sigmaprime}), Kim gets [Eq.~28 in Ref.~\cite{kim1986b}]: 
\begin{equation}
\sigma_r = \frac{1}{4 \pi} \sqrt{\lambda L} 
\label{eq:kim_sigma}
\end{equation}
or, combining equations (\ref{eq:sigmasigmaprime-gaussianbeam}) and (\ref{eq:kimnew_sigmaprime}) [Eq.~6.37 in Ref.~\cite{kim1989}]: 
\begin{equation}
\sigma_r = \frac{1}{4 \pi} \sqrt{2 \lambda L}.
\label{eq:kimnew_sigma}
\end{equation}

Elleaume~\cite{elleaume} followed a different direction. He did not suppose that the observed approximately Gaussian divergence comes from the Fraunhofer propagation of a Gaussian beam as hypothesized by Kim, but obtains a numerical fit of the calculated radiation expressed as a function of the real space at the source position [equation~(\ref{eq:elleaumetheorysize})] and obtained: 
\begin{equation}
\sigma_r = \frac{2.740}{4 \pi} \sqrt{\lambda L} = 0.218 \sqrt{\lambda L}.
\label{eq:elleaume_sigma}
\end{equation}
The fitted Gaussian for spatial (Eq.~\ref{eq:elleaume_sigma}) and angular (Eq.~\ref{eq:elleaume_sigmaprime}) representations of the UR are not related via Fourier transform [equation~(\ref{eq:sigmasigmaprime-gaussianbeam})], or in other words, their product is not $\lambda/(4 \pi)$ but: 
\begin{equation}
\sigma_r  \sigma_{r'} = \frac{1.89 \lambda}{4 \pi} \sim \frac{\lambda}{2 \pi},
\label{elleaume_sigmasigmaprime}
\end{equation}
which can be interpreted as the phase space volume of UR is approximately twice the phase space volume of the first coherent mode of a Gaussian beam. 

In literature, we can find papers that follow Elleaume's [e.g., \cite{borland2012,hettel2014}] and Kim's model [e.g., \cite{huang2013}]. 
To conclude this section, it is worth mentioning, citing \cite{elleaume}, that ``these are approximations and should not be considered as fundamental results". 
Moreover, Walker \cite{Walker2019} affirms that ``The reason why different models for the source size and divergence have been put forward is that the radiation phase space is not at all Gaussian in nature". 
Recently, several papers discussed the undulator's phase space and obtain expressions of brighness, coherence, etc. adapted to new generation of sources \cite{geloni2008, Tanaka2014, lindberg2015, Walker2019}.

In the SHADOW4 code, and for the following discussion, we adopt the Elleaume's approach [Eqs. (\ref{eq:elleaume_sigmaprime}) and (\ref{eq:elleaume_sigma})].

\subsection{Description of electron sizes and emittance}
\label{sec:emittance}

At any position $s$ in the storage ring, an electron can be described by 5 coordinates:
$\mathcal{S}= (x,x',y,  y',\delta_\mathcal{E})$ representing the phase space coordinates and a term $\delta_\mathcal{E}$ expressing the relative deviation of the electron energy from main storage ring energy (also known as the energy spread). It follows that at any given $s$ the many electrons in a bunch follow a 5D Gaussian distribution:
\begin{equation}\label{eq:f-electrons}
f(\mathcal{S}) = \frac{1}{(2 \pi)^{5/2} \sqrt{\text{det}(M)}} \exp
        \left( -\frac{1}{2} \mathcal{S}^\text{T} M^{-1} \mathcal{S} \right),
\end{equation}
with $M$ the generalized variance 5$\times$5 matrix.
A common assumption is that the variables are correlated only if they are in the same plane ($x$ or $y$), thus defining 2$\times$2 matrices. For $x$ (horizontal plane)
\begin{equation}\label{eq:twiss}
    M_{x} = 
\begin{pmatrix}
\langle xx \rangle & \langle x x' \rangle\\
\langle x x' \rangle & \langle x'x' \rangle 
\end{pmatrix}
=
\begin{pmatrix}
\sigma_x^2 & \rho\sigma_x\sigma_{x'}\\
\rho\sigma_x\sigma_{x'} & \sigma_{x'}^2
\end{pmatrix}
= \epsilon_x
\begin{pmatrix}
\beta_x & -\alpha_x\\
-\alpha_x & \gamma_x
\end{pmatrix},
\end{equation}
and similarly for the $y$ coordinate (vertical plane).
We also indicate the expression as a function of the Twiss functions ($\alpha$, $\beta$ and $\gamma$) and emittance $\epsilon_x=(\langle xx \rangle) \langle x' x' \rangle - 2 \langle x x' \rangle)^{1/2}$.
In some particular points or the storage ring, the covariance between spatial and angle terms is zero ($\rho$=$\alpha$=0), thus only the diagonal terms ($\sigma_x^2,\sigma_{x'}^2,\sigma_y^2,\sigma_{y'}^2,\delta_\mathcal{E}^2)$ are sufficient to define the electron beam. This is the case at the centre of the straight sections, where the undulators are usually placed. When the undulator is in another position at a distance $s$ from the center of the straight section (with Twiss parameters $\alpha_0$, $\beta_0$ and $\gamma_0$) the new parameters are: 

\begin{eqnarray}
    \beta_1 = \beta_0 - 2 s \alpha_0 + \gamma_0 s^2, \nonumber \\
    \alpha_1 = \alpha_0 - \gamma_0 s, \\
    \gamma_1 = \gamma_0. \nonumber
\end{eqnarray}
\subsection{Divergence and size of the photon source at resonance in Gaussian approximation}
\label{sec:electronSpread}
Consider a filament beam that emits a radiation wavefront. At resonance, the size and divergence distributions are supposed Gaussians given by equations~(\ref{eq:elleaume_sigma}) and (\ref{eq:elleaume_sigmaprime}), respectively. Consider now that the emission is not given by a filament beam, but instead by a bunch of electrons distributed, as discussed, with values 
$\sigma_x, \sigma_{x'},\sigma_y,\sigma_{y'},\delta_\mathcal{E}$.

Suppose first that all the electrons have exactly the same energy ($\delta_\mathcal{E}$=0).
We can assume that the emission of the photon source is the convolution of the photon source (filament beam) with the electron beam. Therefore, the sizes and divergences of the photon source are:
\begin{subequations}
\begin{align}
   \Sigma_{x,y} &= \sqrt{\sigma_r^2 + \sigma_{x,y}^2}\\
   \Sigma_{x', y'} &= \sqrt{\sigma_{r'}^2 + \sigma_{x',y'}^2}.
\end{align}\label{eq:Convolution}
\end{subequations}

Tanaka and Kitamura~\cite{Tanaka2009} have studied the effect of the electron energy dispersion in the UR. They found that the approximated photon source size and divergences are
\begin{subequations}\label{eq:ConvolutionSpread}
\begin{align}
   \Sigma_{x,y} &= \sqrt{(Q_s \sigma_r)^2 + \sigma_{x,y}^2}\\
   \Sigma_{x', y'} &= \sqrt{(Q_a \sigma_{r'})^2 + \sigma_{x',y'}^2},
\end{align}
\end{subequations}
where corrective terms for sizes $Q_s$ and angles $Q_a$ have been introduced. They depend on the electron ``normalized energy spread"  $\sigma_\epsilon = 2 \pi n N \delta_\mathcal{E}$, with $\delta_\mathcal{E}$ the electron energy dispersion, $N$ the number of periods of the undulator and $n$ the harmonic number in use.
The $Q_a$ functions is 
\begin{equation}\label{eq:Qa}
Q_a(x) = \left[
\frac
{2 x^2}
{-1+\exp(-2 x^2) + \sqrt{2 \pi}~x~\text{erf}(\sqrt{2} x)}
\right]^{1/2},
\end{equation}
with $\text{erf}(x)$ the error function \cite{wiki_erf};
and the $Q_s$ function
\footnote{Note that \cite{Tanaka2009} have an additional factor 2. This is related to the mentioned discussion on calculating the size using equations~(\ref{eq:kim_sigma}) or (\ref{eq:kimnew_sigma}). In our case, it is included in our equation~(\ref{eq:elleaume_sigma}), therefore $Q_s \rightarrow 1$ for zero energy spread $x \rightarrow 0$.} is
\begin{equation}\label{Qs}
    Q_s = [Q_a(x/4)]^{2/3}.
\end{equation}

In \cite{Geloni2018} the authors treat how the brightness is influenced by the energy spread in a context not restricted to Gaussian approximations. They obtain the approximated expressions compatible with those of \cite{Tanaka2009} and argue that they ``may constitute a good approximation in some region of the parameter space, when it comes to the limit for a diffraction-limited beam with non-negligible energy spread, a more detailed study is needed". 


For the calculations in this paper we use the parameters of the ESRF U18 undulator installed in the ID06 beamline at EBS-ESRF. It has a period of $\lambda_u$=\SI{18}{\milli\meter}, $N$=111, therefore length of 2~m.
The $K$ value ranges from 0.2 to 1.479. The resonance is set to $E$=\SI{10}{\kilo\eV} with $K$=1.3411.
The EBS storage ring has electron beam energy $\mathcal{E}$=\SI{6}{\giga\eV}, electron energy spread of $\delta_\mathcal{E}$=\SI{9.334e-4}{} and electron sizes at the center of the straight section where the undulator is installed 
$\sigma_{x}$=\SI{30.18}{\micro\m}, $
\sigma_{y}$=\SI{3.64}{\micro\m}, $
\sigma_{x'}$=\SI{4.37}{\micro\radian}, $
\sigma_{y'}$=\SI{1.37}{\micro\radian}. 
They give beam emittances:  
$\varepsilon_x\text{=}\sigma_x \sigma_{x'}$=\SI{132}{\pico\meter \radian}, and 
$\varepsilon_y \text{=} \sigma_y \sigma_{y'}$=\SI{5}{\pico\meter \radian}.

\begin{figure}
  \centering

\includegraphics[width=1.0\textwidth]{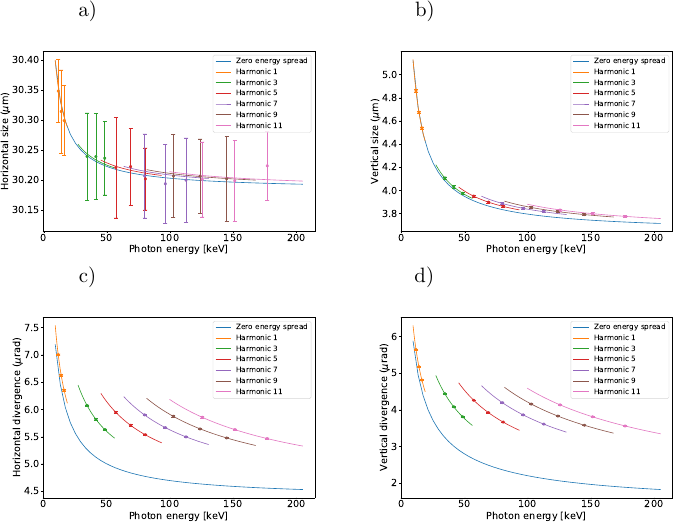}

  \caption{Full width at half-maximum values of size [in the horizontal (a) and vertical (b) directions]  and divergence [in the horizontal (c) and vertical (d) directions] of the photon source as a function of the resonance photon energy for the ID06-U18 at EBS/ESRF.
  The continuous lines are calculated using equations~(\ref{eq:ConvolutionSpread}). 
  The points (dots) corresponding to ray tracing using the ``Undulator Gaussian" widget. Every point corresponds to the average value after 50 SHADOW4 simulations, using a random Monte-Carlo seed each. The corresponding standard deviation is shown as error bars (note the small range of horizontal size in left (a) resulting in large error bars). 
  }
  \label{fig:gauss_size_and_divergence}
\end{figure}

As an example, the undulator sizes and divergences for the ESRF ID06 U18 undulator calculated using equations (\ref{eq:ConvolutionSpread}) are shown in the solid lines of Fig.~\ref{fig:gauss_size_and_divergence}.
From Figs.~\ref{fig:gauss_size_and_divergence}c and \ref{fig:gauss_size_and_divergence}d it is evident that for high harmonics the divergence is influenced by the electron energy spread, thus the necessity of including it in the ray tracing simulations.
The effect is beam size is moderate (Figs.~\ref{fig:gauss_size_and_divergence}a and \ref{fig:gauss_size_and_divergence}b).
In \cite{Walker2019} the sizes and divergences are discussed ({\it ibid.}, Fig.~8,9) in a context of numeric calculations of brightness also including the energy spread. The results for divergence r.m.s. agree well with the model used here. For the size r.m.s. they observe a discrepancy mainly due to changes in the shape (narrowing the cone and introducing wide tails) of the distribution that becomes less and less Gaussian when increasing $\delta_\mathcal{E}$. 


\begin{figure}
  \centering  
  a)~~~~~~~~~~~~~~~~~~~~~~~~~~~~~~~~~~~~~~~~~~~~~~b)~~~~~~~~~~~~~~~~~~~~~~~~~~~~~\\
  \includegraphics[width=0.49\textwidth]{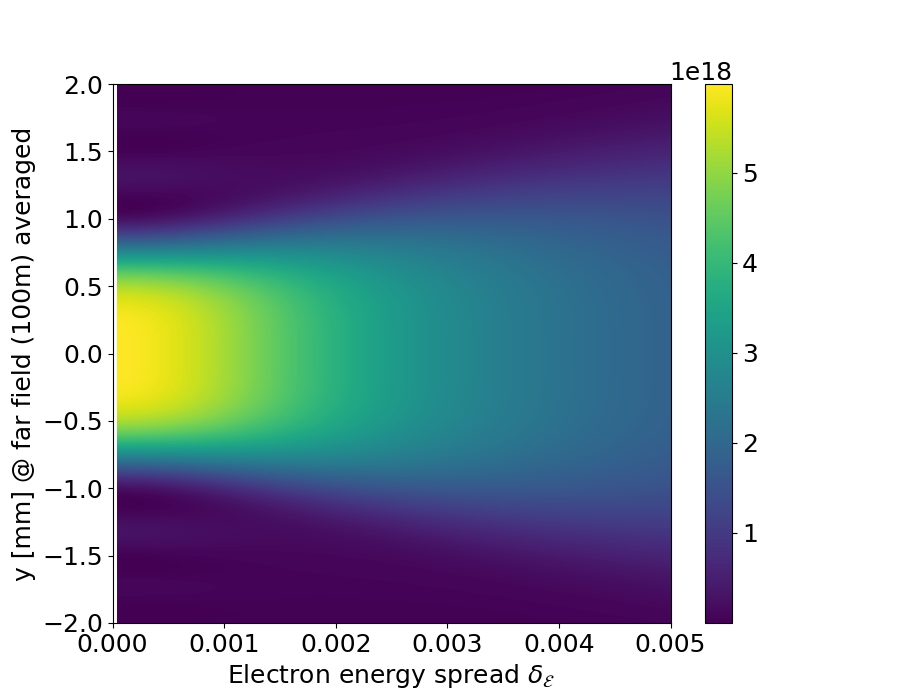}
 \includegraphics[width=0.49\textwidth]{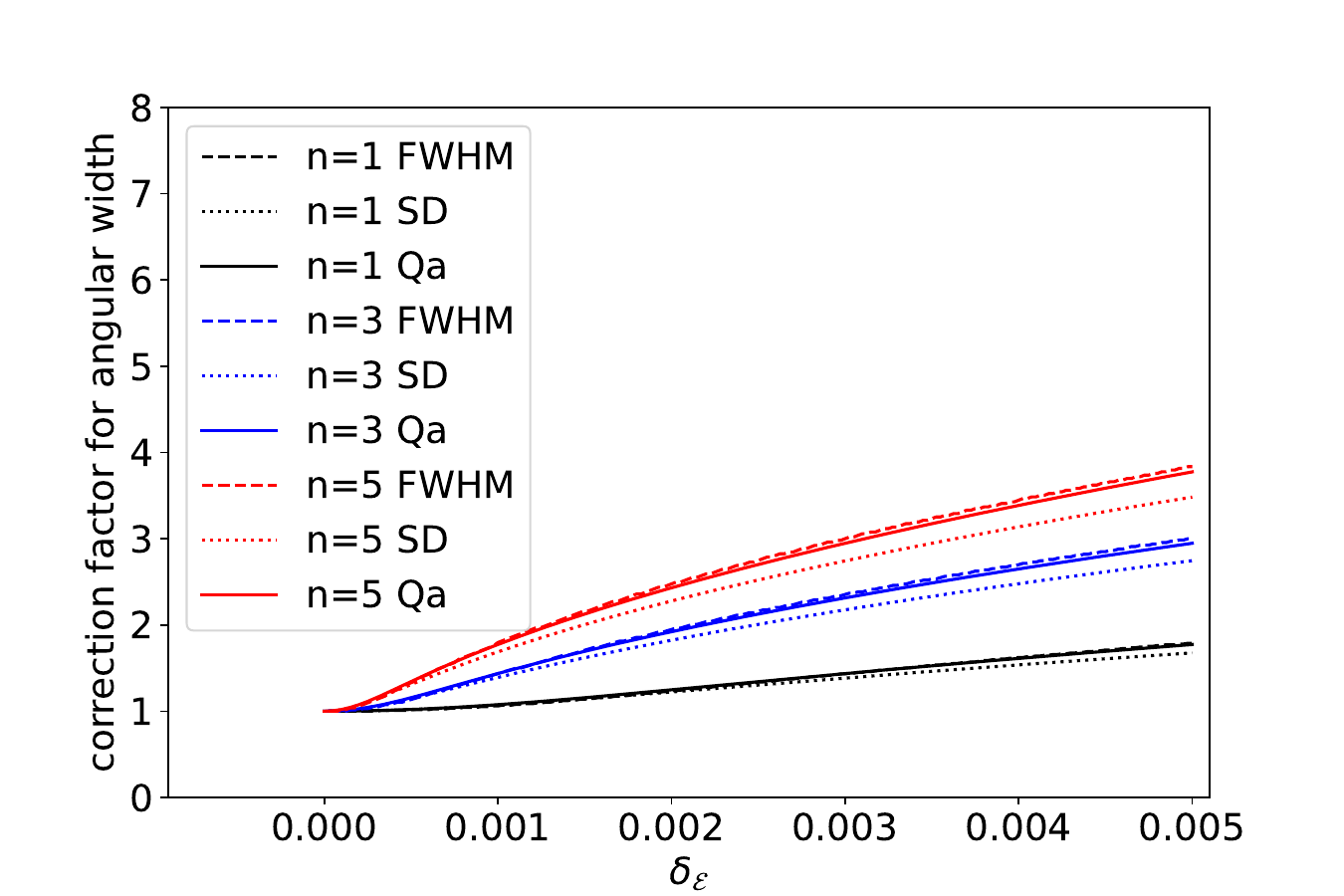}
  c)~~~~~~~~~~~~~~~~~~~~~~~~~~~~~~~~~~~~~~~~~~~~~~d)~~~~~~~~~~~~~~~~~~~~~~~~~~~~~\\
  \includegraphics[width=0.49\textwidth]{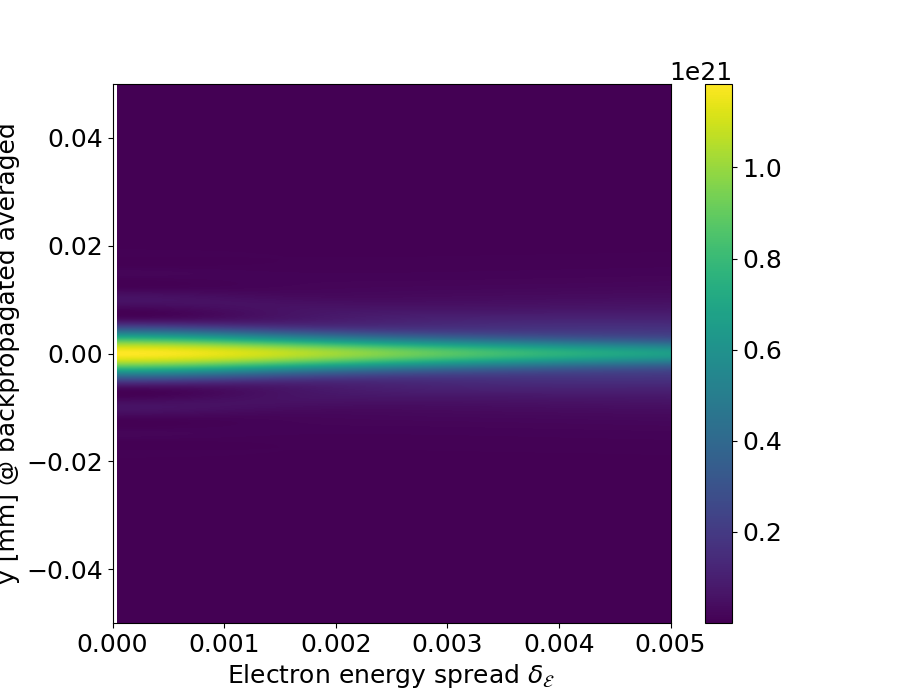}
 \includegraphics[width=0.49\textwidth]{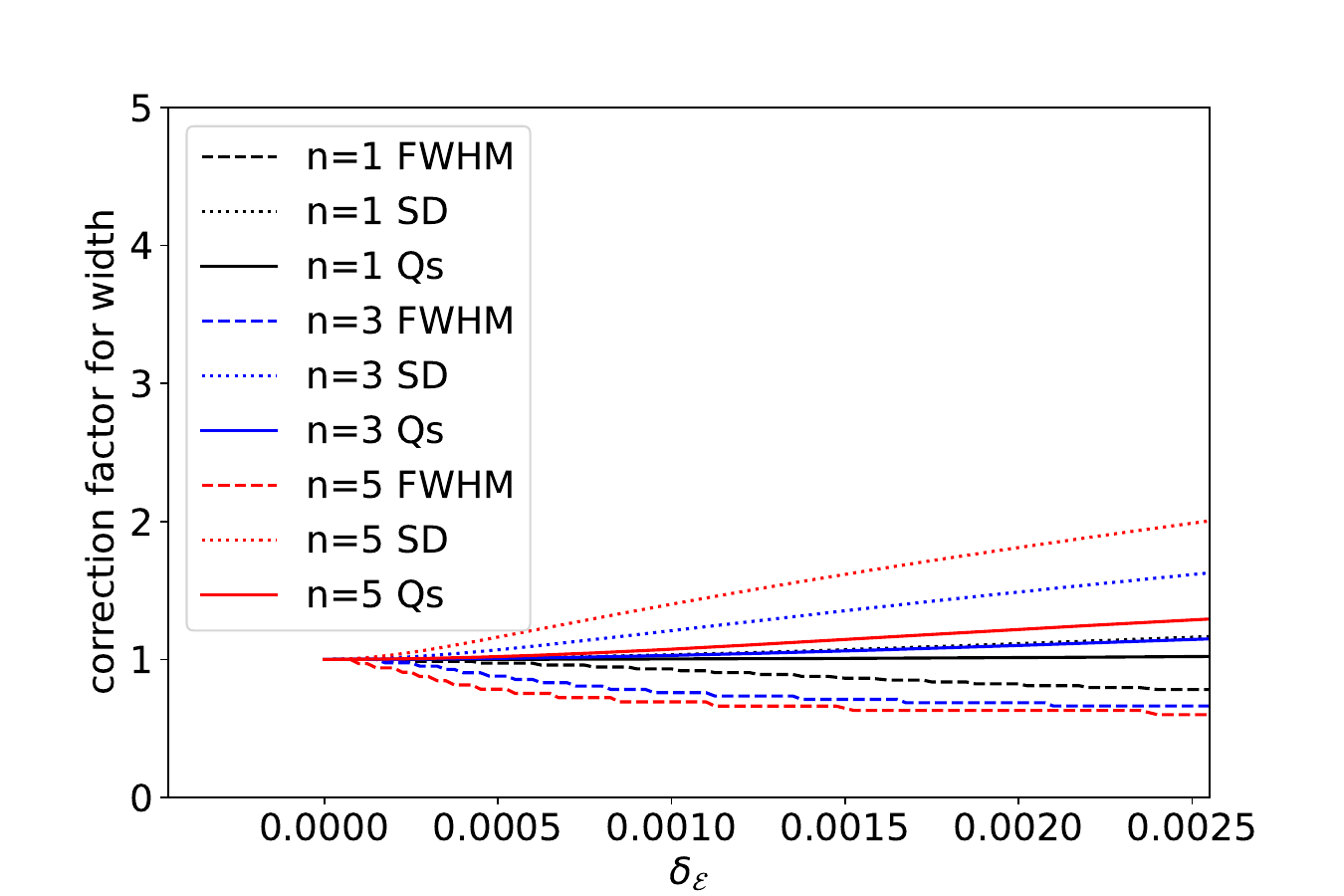}
  \caption{
  Results of wave optics simulation of the intensity versus electron energy spread calculated for ID06 U18 undulator. The radiations is calculated at far field (a and b) and backpropagated to the center of the undulator (c and d). 
  a) intensity map at far field as a function of $\delta_\mathcal{E}$ calculated at the first harmonic (\SI{10}{\kilo\eV}).
  b) Values of the angular-width correction (normalized to the width at zero energy dispersion) as a function of the electron energy spread. The widths are measured by the FWHM and the SD value (2.355$\times$r.m.s.) of the far field intensity profiles (as shown in (a) for $n$=1). They are compared with $Q_a$.
  c) the same as (a) for the backpropagated radiation.
  d) the same as (b) but for the size corrections compared with $Q_s$.
  }
  \label{fig:angular_width_vs_spread}
\end{figure}

We verified numerically the suitability of $Q_a$ to correct the angular width of the undulator emission for the previously discussed ESRF ID06 U18 undulator tuned at \SI{10}{\kilo\eV} (first harmonic).
We supposed here zero emittance and a variable energy spread $\delta_\mathcal{E}$ from zero to 0.005. 
We calculated numerically, using the WOFRY wave optics package \cite{codeWOFRY} 
the intensity distribution at \SI{100}{\meter} (far field).
Without changing the undulator configuration, we repeated the calculation for different values of electron energy around $\mathcal{E}$=\SI{6}{\giga\eV}. 
We then constructed the pattern for each energy spread by doing the sum of the patterns for each electron energy weighted by a Gaussian with the corresponding $\delta_\mathcal{E}$.
We finally calculated the FWHM and the r.m.s. values of each intensity pattern and normalize them to the value obtained for $\delta_\mathcal{E}$ = 0. 
Figure~\ref{fig:angular_width_vs_spread} shows the results calculated for the far field (\ref{fig:angular_width_vs_spread}a and \ref{fig:angular_width_vs_spread}b) and backpropagated to the center of the ID (\ref{fig:angular_width_vs_spread}c and \ref{fig:angular_width_vs_spread}d).
In Fig.~\ref{fig:angular_width_vs_spread}b it is shown the comparison of $Q_a$ with the numerical values at far field of the FWHM and SD (a width calculated from the r.m.s. as if it was Gaussian, i.e.,  SD=2.355$\times$r.m.s.). 
We observe a good agreement of $Q_a$ with the numerical values (both FWHM or SD) for $n$ = 1. The agreement is less good for higher $n$ and $\delta_\mathcal{E}$.
For the backpropagated radiation we see, as previously noticed, that the effect of the electron energy spread is moderate. Indeed, it seems from Fig.~\ref{fig:angular_width_vs_spread}c that there is a shrink in the width when increasing $\delta_\mathcal{E}$.
When examining the numerical values in Fig.~\ref{fig:angular_width_vs_spread}d we observe a discrepancy between the FWHM and SD, indicating a non-Gaussian behaviour.
While the SD increases (as predicted by $Q_s$), the observed FWHM slightly decreases.
This is attributed to a narrowing of the peak accompanied by an expansion of the tails, as noted by \cite{Walker2019}.
In summary, the positive takeaway is that, in all cases, $Q_a$ and $Q_s$ values fall between the numerical values of FWHM and SD, highlighting the difficulty of selecting a single parameter to describe a non-Gaussian distribution.
Keeping in mind that the value of $\delta_\mathcal{E}$ is close to 0.001 for most synchrotron sources, we conclude that the use $Q_a$ and $Q_s$ is a reasonable choice for incorporating the electron energy dispersion in ray tracing simulations when working at resonance.
\subsection{Divergence and size of the photon source off-resonance}
\label{sec:offresonance}

\begin{figure}
  \centering 
  a)~~~~~~~~~~~~~~~~~~~~~~~~~~~~~~~~~~~~~~~~~~~~~~b)~~~~~~~~~~~~~~~~~~~~~~~~~~~~~\\\includegraphics[width=0.49\textwidth]{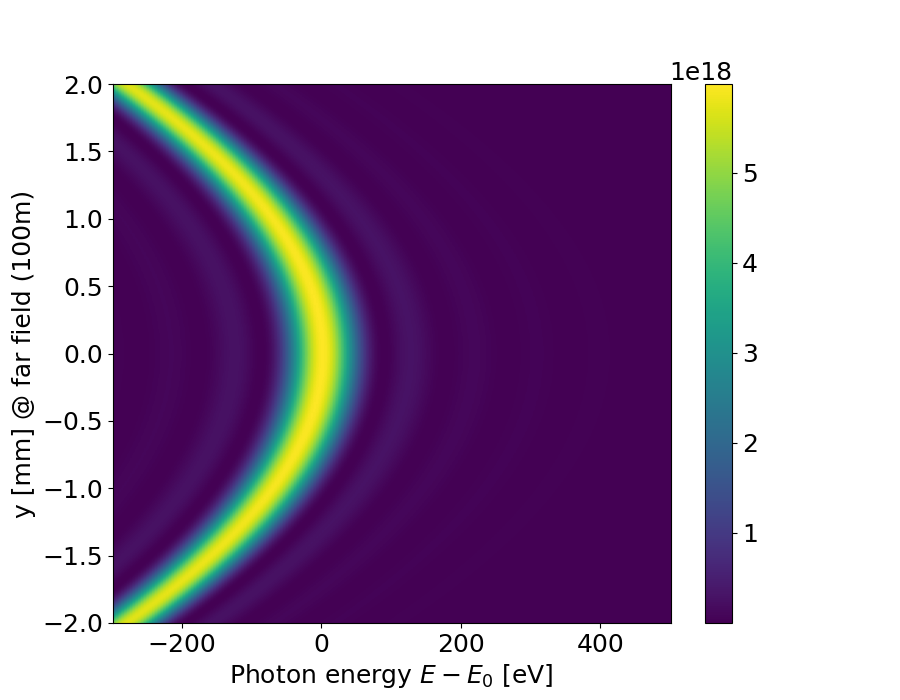}
  \includegraphics[width=0.49\textwidth]{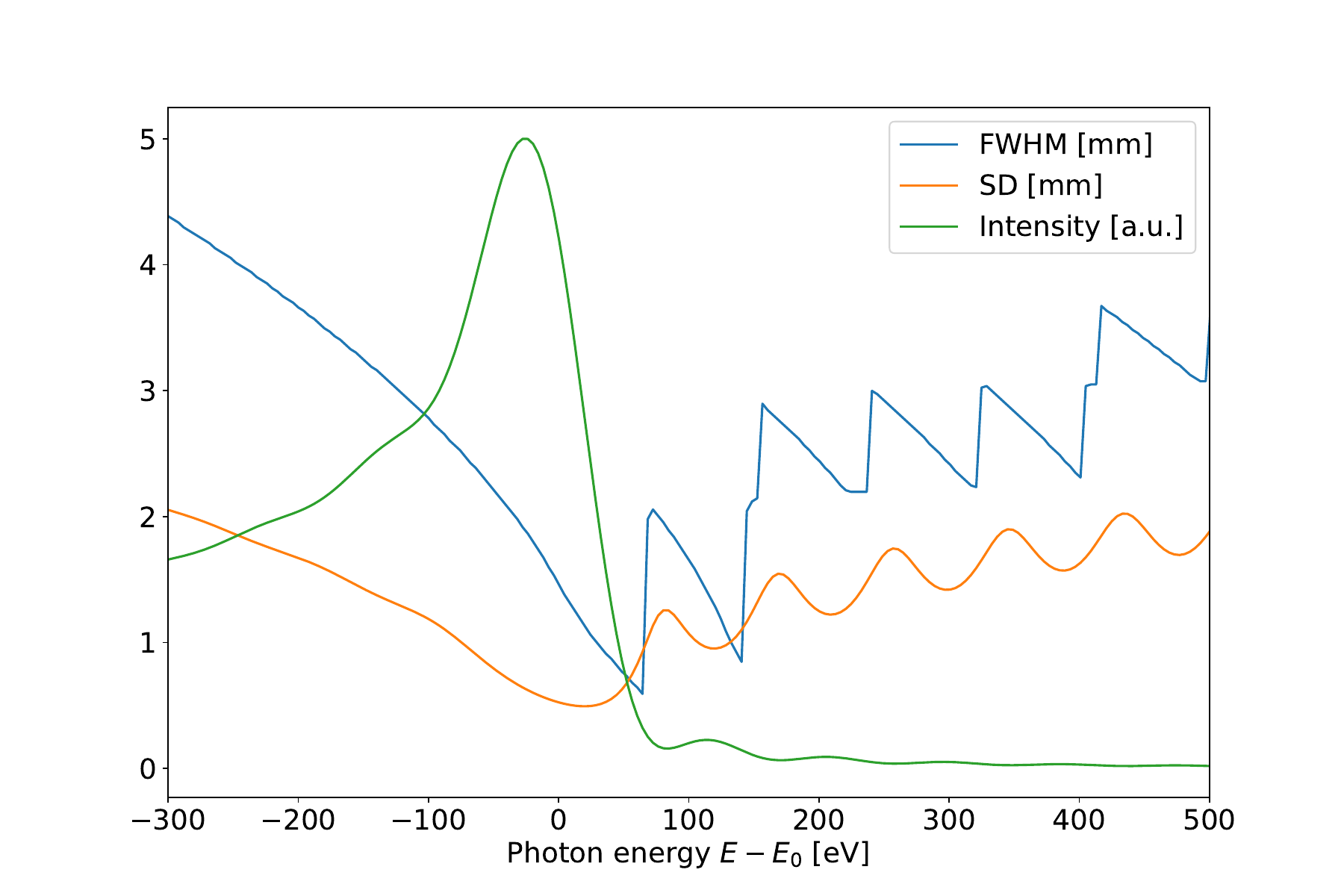}
  c)~~~~~~~~~~~~~~~~~~~~~~~~~~~~~~~~~~~~~~~~~~~~~~d)~~~~~~~~~~~~~~~~~~~~~~~~~~~~~\\
\includegraphics[width=0.49\textwidth]{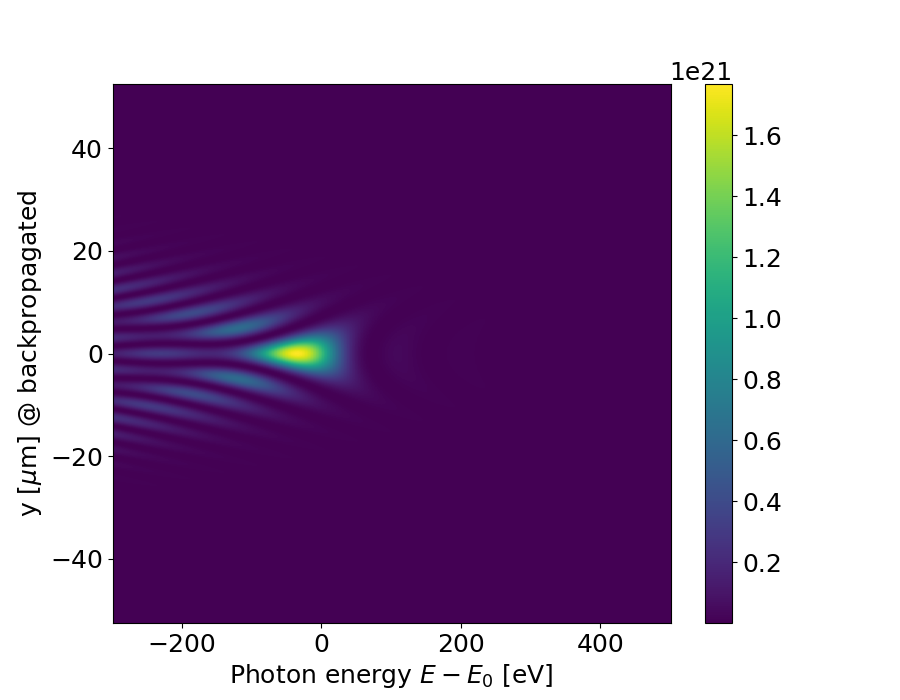}
  \includegraphics[width=0.49\textwidth]{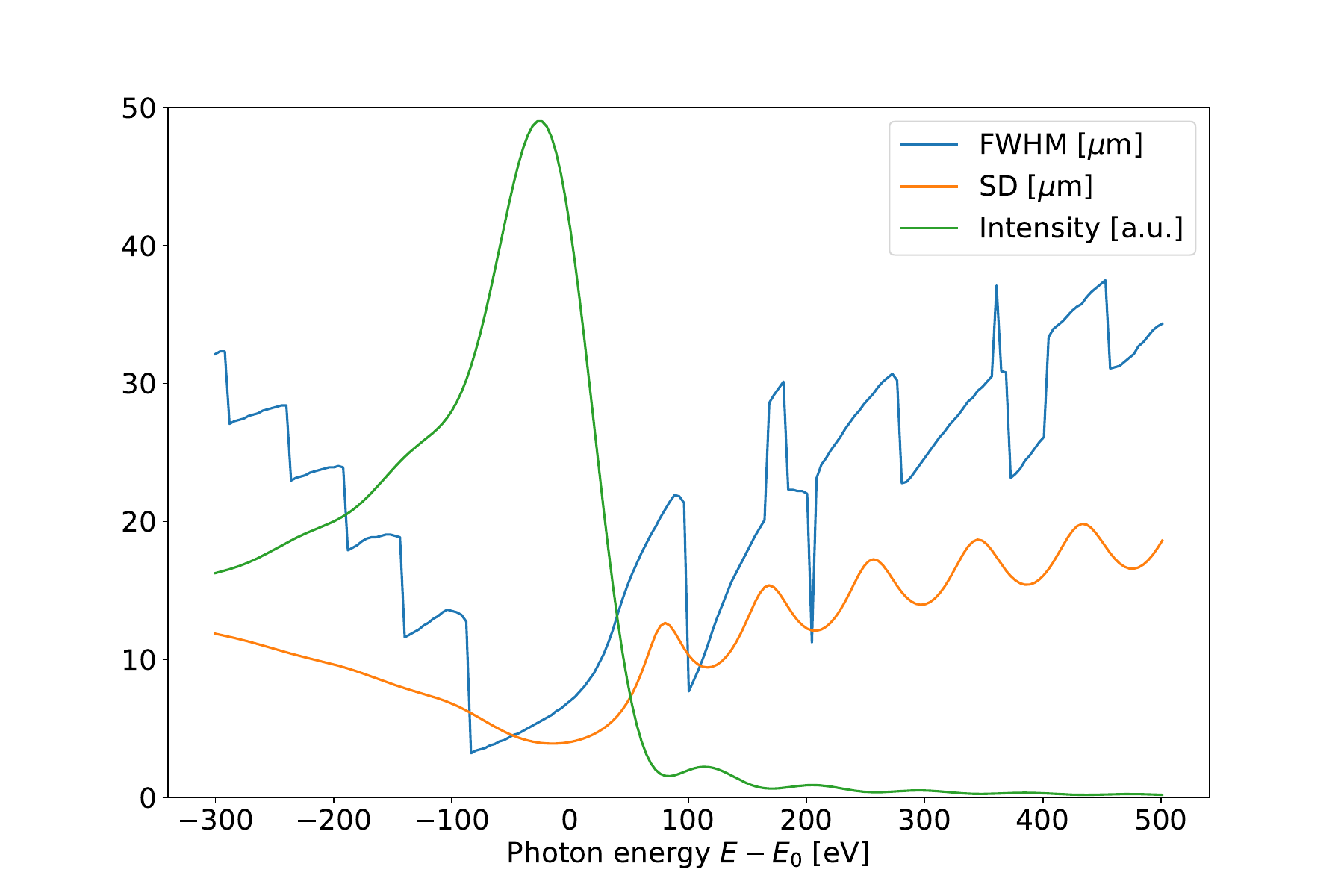}
  
  \caption{
  Results of wave optics simulation of the intensity versus photon energy calculated for ID06 U18 undulator. The radiations is calculated at far field (a and b) and backpropagated to the center of the undulator (c and d). 
  a) Intensity map at far field (100~m) as a function of photon energy calculated for the first harmonic ($E_0$=\SI{10}{\kilo\eV}).
  b) Values of the width of the intensity distribution at the far field as a function of the photon energy.
  The widths are obtained from the FWHM and the SD value (2.355$\times$r.m.s.).
  The intensity (in arbitrary units) is also shown (green curve).
  c) the same as a) for the backpropagated radiation at the center of the undularor.
  d) the same as b) for the backpropagated radiation at the center of the undularor.}
  \label{fig:detuning}
\end{figure}

The values of beam size and divergence obtained in the previous section are valid only when working with photons at resonance (or at a particular odd harmonic). It is common that the experimentalist set the monochromator close, but not exactly at resonance. For example, the photon energy corresponding to the maximum intensity integrated over a finite $\theta$ interval (e.g. using a slit) is not exactly at resonance, but red-shifted by an amount that depends on the aperture.
Moreover, going out of resonance, the intensity distribution changes from a well defined peak (Fig.~\ref{fig:undulator_fit}) to other shapes, also presenting a double-peak. 
This is illustrated in Fig.~\ref{fig:detuning} where numeric values of FWHM and SD are computed for photon energies scanning the first harmonic peak.
It can be observed (see Fig.~\ref{fig:detuning}b) that the minimum of the divergence is obtained at a position blue-shifted with respect to the resonance, but ( see Fig.~\ref{fig:detuning}d) the minimum FWHM of the size tends to a red-shifted position. 

\begin{figure}
  \centering
  a)~~~~~~~~~~~~~~~~~~~~~~~~~~~~~~~~~~~~~~~~~~~~~~~b)~~~~~~~~~~~~~~~~~~~~~~~~~~~~~~~~~~~~~~~~~~~~~\\
\includegraphics[width=0.49\textwidth]{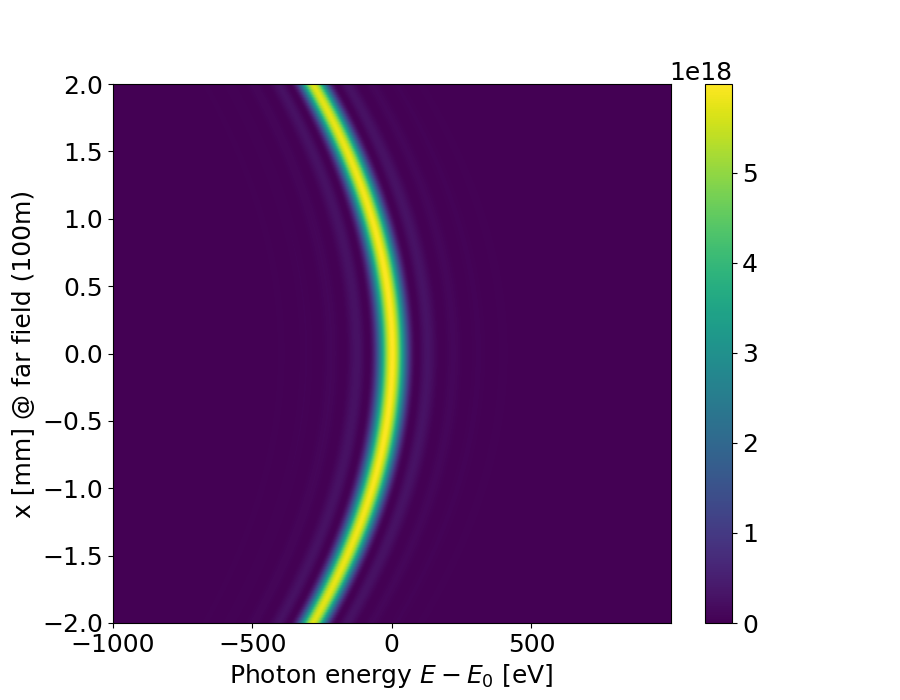}
  \includegraphics[width=0.49\textwidth]{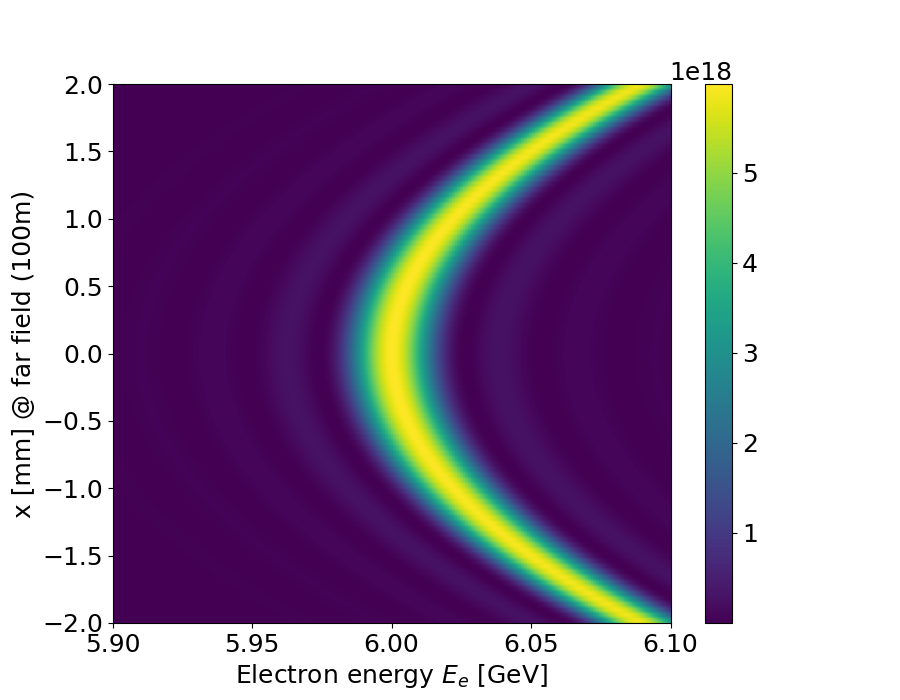}

  \caption{Maps of the intensity at far field (at 100m from the source, $x$ is the vertical spatial coordinate). 
  a) intensity vs photon energy shift from resonance ($E_0$=\SI{10}{\kilo\eV}) and $x$ ($\mathcal{E}$=\SI{6}{\giga\eV});  b) intensity vs electron energy (at resonance $E_0$). 
  }
  \label{fig:detuning-maps}
\end{figure}

The effect of detuning of the electron energy has a similar effect than detuning the photon energy from resonance.
Indeed, from equation~(\ref{eq:resonance}), $\lambda_0 \gamma_0^2\text{ = }(\lambda_u/2)(1+K^2/2)$ when $\theta\text{ = }0$ (on-axis).
Therefore $\lambda_0 \gamma_0^2$ is a constant for a particular tuned undulator, or in other words, a wavelength (or energy) shift is equivalent to a corresponding shift of the  electron energy. One can compensate the other as far as $\gamma_0^2 E_0 = \text{cte}$.
This is illustrated in Figs.~\ref{fig:detuning-maps}a and \ref{fig:detuning-maps}b. 

\begin{figure}
  \centering  
  
  a)~~~~~~~~~~~~~~~~~~~~~~~~~~~~~~~~~~b)~~~~~~~~~~~~~~~~~~~~~~~~~~~~\\
  \includegraphics[width=0.49\textwidth]{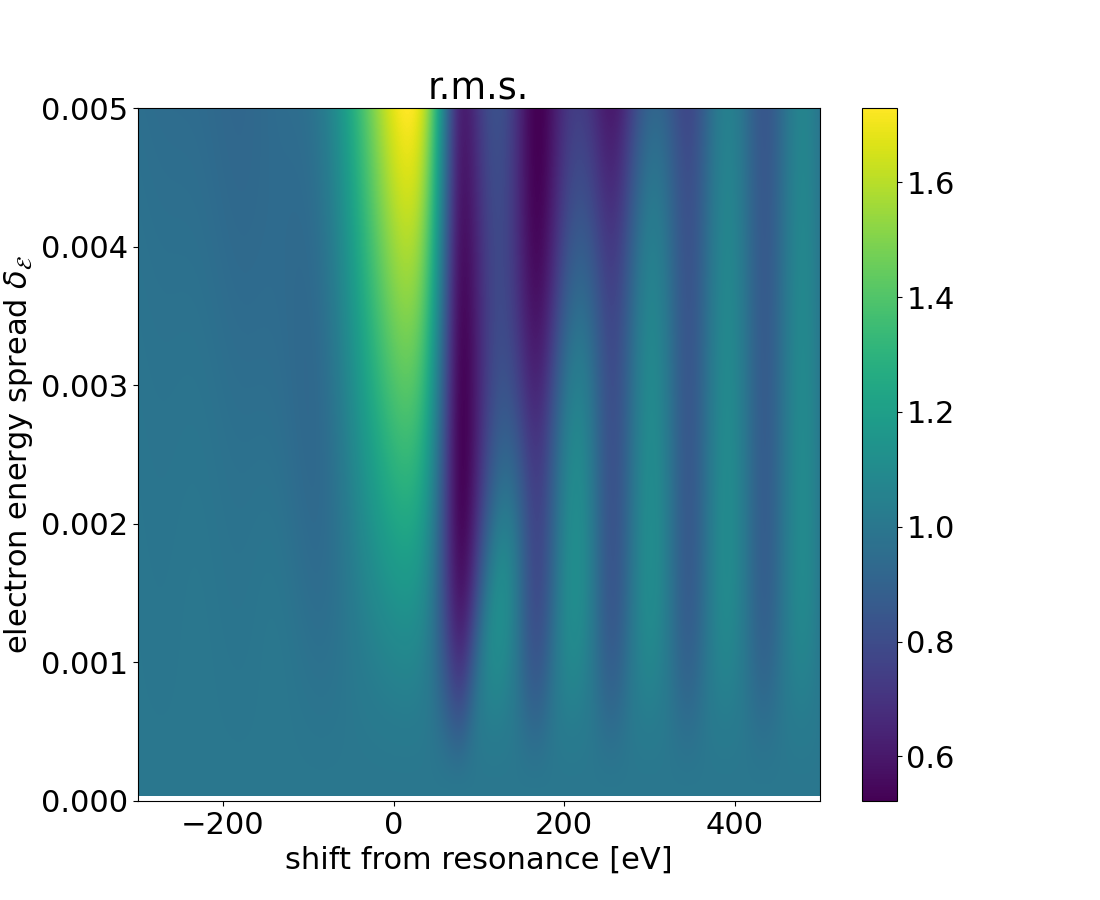}
  \includegraphics[width=0.49\textwidth]{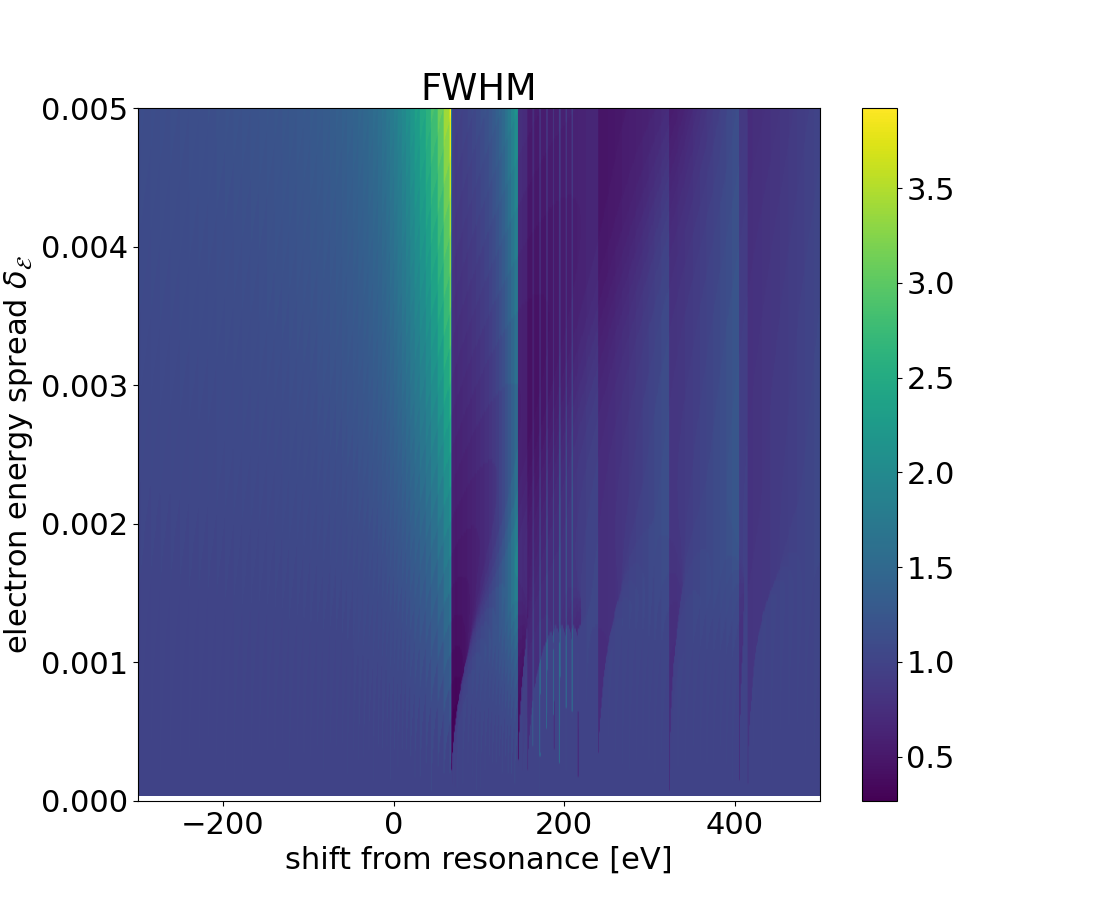}

  c)~~~~~~~~~~~~~~~~~~~~~~~~~~~~~~~~~~d)~~~~~~~~~~~~~~~~~~~~~~~~~~~~\\
  \includegraphics[width=0.49\textwidth]{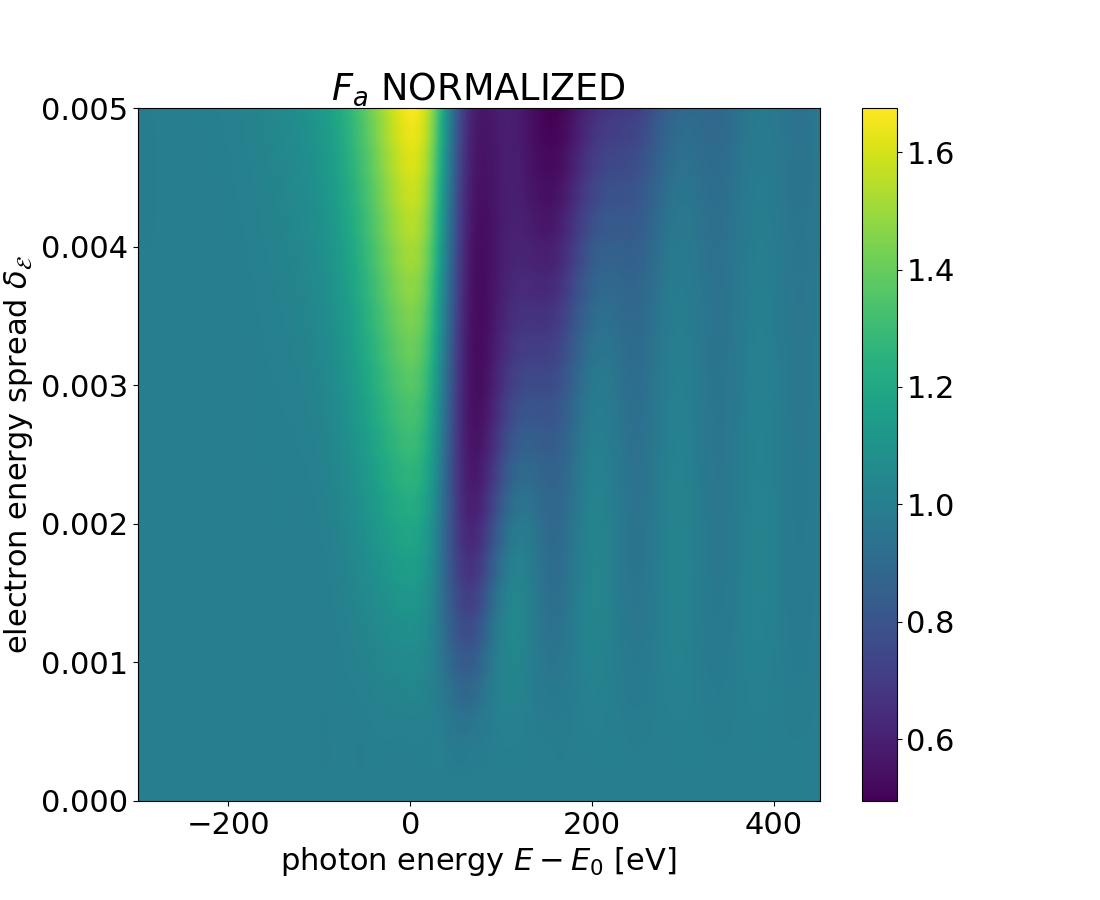}
  \includegraphics[width=0.49\textwidth]{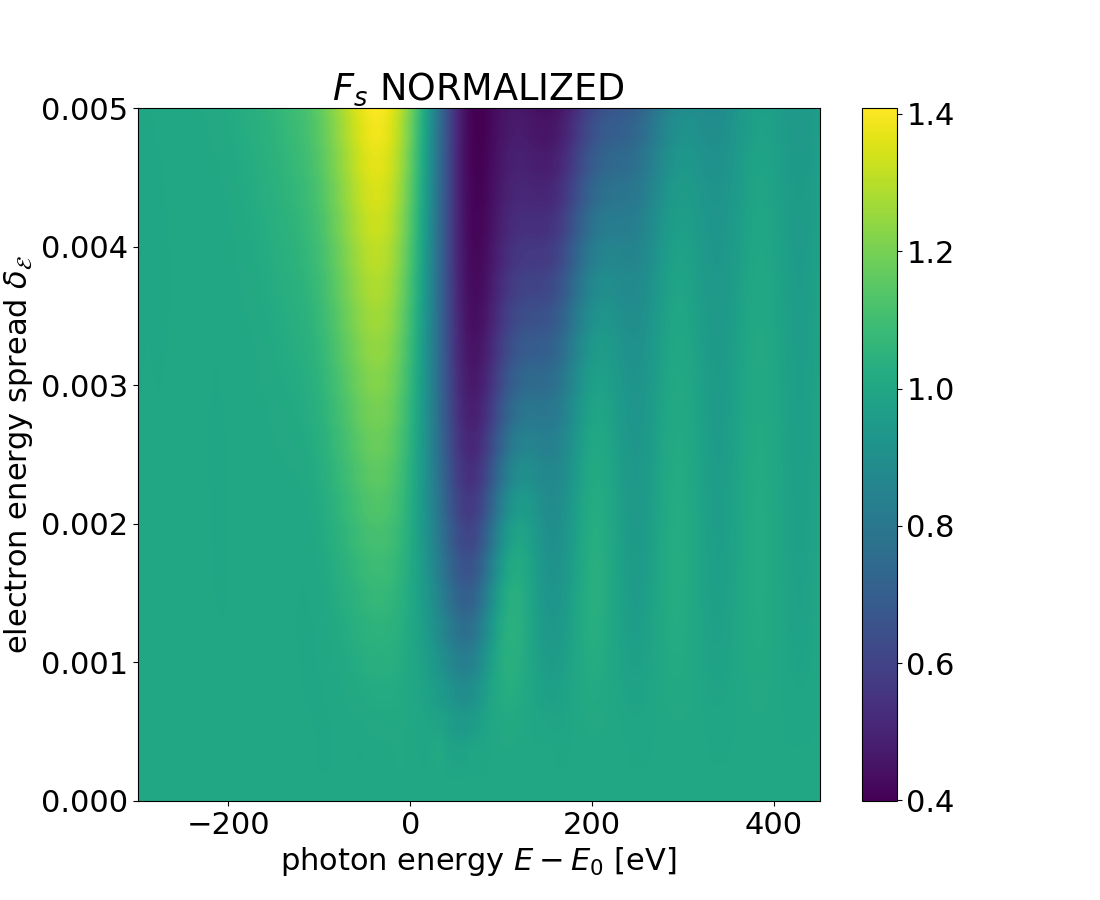}

  \caption{
  a) Calculation of the r.m.s. value of the intensity distribution at 100 m  downstream ID06 U18 undulator.
  b) The same but the map represents the FWHM.
  c) The correction function $F_a$ calculated using data from \cite{Nash2019}] applied to the same undulator. 
  d) The correction functions $F_s$ calculated using data from \cite{Nash2019}] applied to the same undulator.
  Data in each maps are normalized the values at zero electron energy spread.}
  \label{fig:universalnormalizedfunctions}
\end{figure}

In \cite{Nash2019} the authors present functions to correct the flux, size, and divergence considering the electron energy spread and the detuning from the resonance. 
They built 2-dimensional maps of functions that correct the flux, size and divergence standard deviation, versus both the energy spread and the spectral detuning.
In this way, the mentioned correction functions $Q_{a,s}(\sigma_\epsilon)$ are replaced by functions $F_{a,s}(E-E_0, \sigma_\epsilon)$ also including the detuning from resonance $E-E_0$. 
Following the same idea, we calculated numerically using WOFRY the maps of the r.m.s. (Fig.~\ref{fig:universalnormalizedfunctions}a) and  the FWHM (Fig.~\ref{fig:universalnormalizedfunctions}b) of the far field intensity for the ESRF ID06 undulator around the resonance.
Because we want to observe the changes versus $\delta_\mathcal{E}$ (the dependency on $E-E_0$ is discussed in the last section), we normalize each value to the corresponding value at $\delta_\mathcal{E}$=0.
The differences observed between these two maps indicate, again, that the distributions are not Gaussian therefore FWHM is not a related to r.m.s. by the 2.355 constant. 
Another observation is that the most changes (values that separate from one) are observed for small values of $E-E_0$. 
We compared these results with the values from \cite{Nash2019} applied to out particular undulator. The map for angles (Fig.~\ref{fig:universalnormalizedfunctions}c) agrees well with our corresponding r.m.s. map (Fig.~\ref{fig:universalnormalizedfunctions}a). For completeness we also applied the on size at source position to our undulator Fig.~\ref{fig:universalnormalizedfunctions}d).
Our results, confirmed by those of  Nash {\it et al.} \cite{Nash2019}, conclude that the corrections by energy spread are important at the resonance, but are not so important far from resonance. This remark will be used in the undulator model presented in Section~\ref{sec:fullundulator}.

\section{Algorithms used in ray tracing undulator sources}\label{sec:raytracing}

We present in this section the ideas behind the models for creating an undulator source with SHADOW4, in its two different applications: the simplified and approximated ``Undulator Gaussian", and the more accurate ``Undulator Light Source".

\subsection{The ``Undulator Gaussian"}
\label{sec:gaussianundulator}

It is generally a good idea to start ray tracing simulations for prototyping an undulator beamline with a simplified and quick model.
This is done with the ``Undulator Gaussian". It supposes that we work at the resonance energy, or at energy very close to it to suppose the results at resonance are applicable.
The rays are sampled following Gaussian distributions in both size and divergence.
We use the equations~(\ref{eq:Convolution}) to calculate the sigmas of the photon source.
For this, the only requirement is to know the electron beam sizes, the working photon energy, and the undulator length. In SHADOW4 we give the option to consider the effect of the electron energy spread.
If we activate this option, the user is requested to enter the values of $\delta_\mathcal{E}$ and also $N$ and $n$ (the undulator period and the harmonic in use) to correct the sizes and divergences using equations~(\ref{eq:ConvolutionSpread}).

Due to the assumptions made, the source is considered monochromatic. 
However, when modelling crystal monochromators, which typically have a very narrow energy bandwidth, it is beneficial to generate a polychromatic source over an energy range slightly broader than the monochromator's acceptance.
To achieve this, we introduce an energy interval $\Delta E$ around the resonance, within which ray energies are sampled according to a simplified flat distribution.

A key enhancement in SHADOW4 is that the sources now include information on the number of photons, allowing the application to directly provide data on absorbed and transmitted intensity and power. This eliminates the need to manually rescale the number of rays to represent the number of photons.
In the simplified model of the ``Undulator Gaussian", the user can either enter this value manually or allow it to be calculated using the equation of the flux in the central cone [equation (17) in \cite{KimInXraydatabooklet}]:

\begin{equation}
    F = \pi \alpha N \frac{\Delta\omega}{\omega} \frac{I_{SR}}{e}Q_n(K), n \text{~~odd}
    \label{eq:flux_central_cone}
\end{equation}
where $\alpha$ is the fine-structure constant, $I_{SR}$ is the electron beam current, $\Delta\omega/\omega$ is the photon energy bandwith, typically 10$^{-3}$; and   $Q_n(K)\text{=}F_n(1+K^2/2)(1/n)$, with $F_n$ a universal function defined in \cite{KimInXraydatabooklet}.

\subsection{The ``Undulator Light Source"}
\label{sec:fullundulator}

This is the primary application to simulate undulator sources.
The user selects input parameters for the electron beam (sizes, divergences, energy and current), undulator parameters ($N$, period and $K$) and sampling parameters (number of rays $N_{\text{rays}}$ and the photon energy interval).
An important parameter is $\theta_{max}$, the maximum radial angle to be considered in the calculations. It affects the total number of photons and the sampling of the rays.
Other ``advanced" parameters permit to define the number of sampling points, and flags to activate some modeling options described below.

In SHADOW4, like in the original SHADOW1, the sampling of points that follow a given (1D, 2D or 3D) distribution is done using the ``inverse method", an optimized algorithm proposed by John von Neumann in a famous letter to Stan Ulam [Fig.~3 in \cite{Eckhardt}]. 
The application of this method to sample undulator ray energies and divergences is discussed in detail in Section~6 of \cite{shadow2undulators}. 

The steps to create the SHADOW4 source are the following: 
\begin{itemize}
    \item Construct stack of the electric fields (for $\sigma$ and $\pi$ polarizations) of the radiation emitted by the undulator in the far field as a function of the photon energy, radial angle $\theta$ and azimuthal angle $\varphi$ for the filament beam. This requires a previous calculation of the energy trajectory. This is a 3D stack as of $N_E \times N_{\theta} \times N_{\varphi}$ 
    points, which is set by the user. In the case that the user wants a ``monochromatic" source $N_E$=1. This main step uses the theory of undulators based on equation~(\ref{eq:undulator}). Like in \cite{shadow2undulators} the stack is computed in polar coordinates, that is more efficient than using cartesian coordinates, because the ``almost" axial symmetry permits to limit $N_{\varphi}$ to low values.
    \item From these stacks of electric field for the two polarizations  $A_{\omega,\sigma}$ and $A_{\omega,\pi}$, compute the stacks of intensity $\mathcal{I}\text{ = }|A_{\omega,\sigma}|^2 + |A_{\omega,\pi}|^2$ and polarization $\mathcal{P}\text{ = }|A_{\omega,\sigma}|/(|A_{\omega,\sigma}|+|A_{\omega,\pi}|)$
    \item When the electron energy spread is taken into account, the array containing the angular array $\theta$ is multiplied by the Tanaka's function $Q_a$ calculated at the resonance. This option is only allowed when simulating monochromatic sources. As discussed before, the user should only activate this correction when working at the resonance or very close to it. 
    \item $\mathcal{I}$ is used as a 3D probability density function, thus integrate it  to obtain the 2D and 1D cumulative distribution functions.
    Then obtain the sampled arrays of the photon energy and divergences (directions) of the rays.
    Up to here, the rays are directed as if there were emitted by a filament beam.
    Then, correct these directions for electron beam emittance by adding sampled values that follow Gaussian distributions with $\sigma_{x'}$ and $\sigma_{y'}$. This is equivalent to perform the mathematical convolution. In case of finite Twiss $\alpha$, use a 2D Gaussian as defined in equation~(\ref{eq:twiss}).
    \item Calculate the polarization for each ray by interpolating $\mathcal{P}$ with the sampled $E_i, \theta_i, \varphi_i$ values. With it, construct the SHADOW electric field vectors $\textbf{A}_\sigma$ and $\textbf{A}_\pi$.
    The first is directed along the horizontal axis and the second along the vertical axis. The intensity for each ray is normalized to one:  $|\textbf{A}_\sigma|^2 + |\textbf{A}_\pi|^2 = 1$.
    \item Sample the ray positions. For the undulator, the photon source has ``no depth", i.e., $z$ = 0 for all rays. Contrary to other sources (bending magnets and wigglers) where the rays start from different positions along the electron path, the undulator rays start from a volume corresponding to the backpropagation of the far field to the plane $z$ = 0. The horizontal and vertical distribution are due to i) the backpropagated far emission or the filament beam (a sort of ``diffraction limit" size), and ii) the electron sizes $\sigma_x$ and $\sigma_y$. They are calculated and combined as follows:
    \begin{itemize}
        \item The backpropagated field for the filament field can be selected among three options:
        i) to neglect it, setting a point source $x$=$y$=$z$= 0 (before convolution with electron beam sizes), as done in the first model of SHADOW1 (a good approximation for ``high-emittance" storage rings like those of 20$^\text{th}$ century because the electron sizes are much larger than the ``diffraction limited" sizes);
        ii) sample rays following the Gaussian approximation for the emission at resonance [equations~(\ref{eq:elleaume_sigma})] (this is a good approximation when simulating monochromatic sources at the resonance); and 
        iii) use a most accurate method backpropagating the far field radiation to the plane $z$ = 0 and sampling rays accordingly.
        This is a costly and delicate operation as it involves a careful sampling of the radiation field, and implementing and setting the wave propagation for the $N_E$ photon energies.
        A compromise has been found to get a reasonable solution without exploding the calculation time (see appendix A). 
        \item If the option of considering the electron energy spread is on, the array with the sizes $r$ is multiplied by the Tanaka's correcting function $Q_s$.
        Again this option is only available for monochromatic sources.  
        \item Once the size distribution is found, the rays are sampled and corrected by adding the Gaussian sampling of the electron source with $\sigma_x$ and $\sigma_{x'}$, or from a 2D Gaussian [equation~(\ref{eq:twiss})] in case of finite Twiss $\alpha$. 
    \end{itemize}
    \item The collection of rays with sampled photon energies, directions (divergences), sizes, and electric field (polarization) constitute the ray tracing source.
\end{itemize}

\begin{figure}
  \centering  \includegraphics[width=0.8\textwidth]{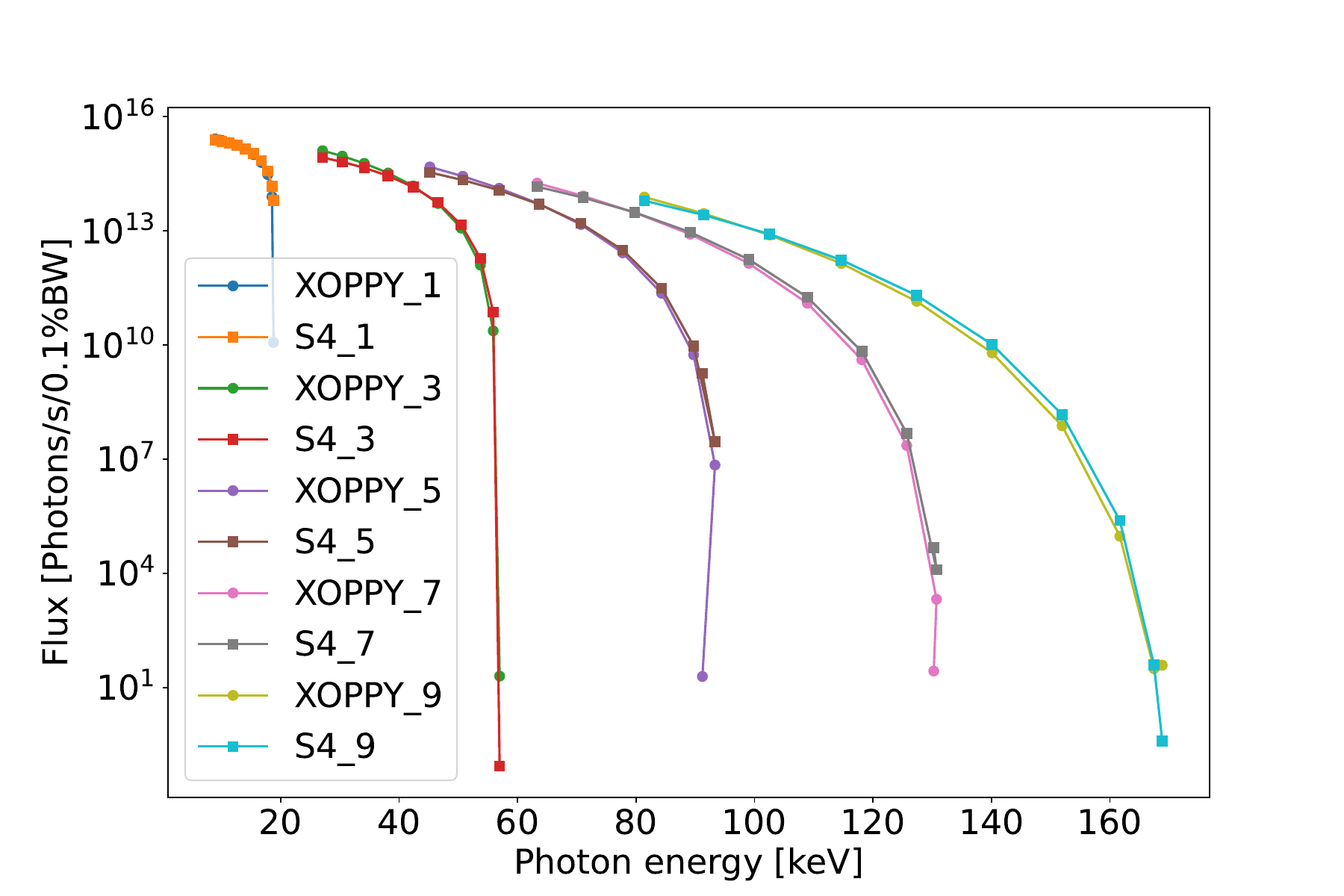}
  \caption{Flux comparison from harmonic 1 to 9 of SHADOW 4 Gaussian undulator and XOPPY/SRW.}
  \label{fig:comparison_tuning_curves}
\end{figure}

\section{Examples and discussion}
\label{sec:examples}

Here, we present calculations using the two SHADOW4 undulator sources, the ``Undulator Gaussian" and the ``Undulator Light Source" that implement the methods described in sections ~\ref{sec:gaussianundulator} and \ref{sec:fullundulator}, respectively. 
The aim is to confirm that the rays generated at the source to represent the undulator accurately match the expected intensity distributions based on theoretical predictions. This serves as a benchmarking process to validate the reliability of the new code.

We use for the calculation the ESRF U18 undulator described before. 
For testing the the ``Undulator Gaussian", several simulations have been performed in monochromatic mode at different photon energies. From the intensity distributions as a function of spatial or angular coordinates the FWHM has been evaluated. 
The results are shown in figure (\ref{fig:gauss_size_and_divergence}),  overplotted to the calculations using the analytical expressions of the energy spread effect [equations (\ref{eq:ConvolutionSpread})].
The plot includes error bars in the SHADOW simulations, derived from the standard deviation of multiple runs. 
It is observed that the theoretical value falls within the error bars of the ray tracing results, confirming that the parameters extracted from the rays do align with the underlying theoretical model.


\begin{figure}
  \centering
\includegraphics[width=1.0\textwidth]{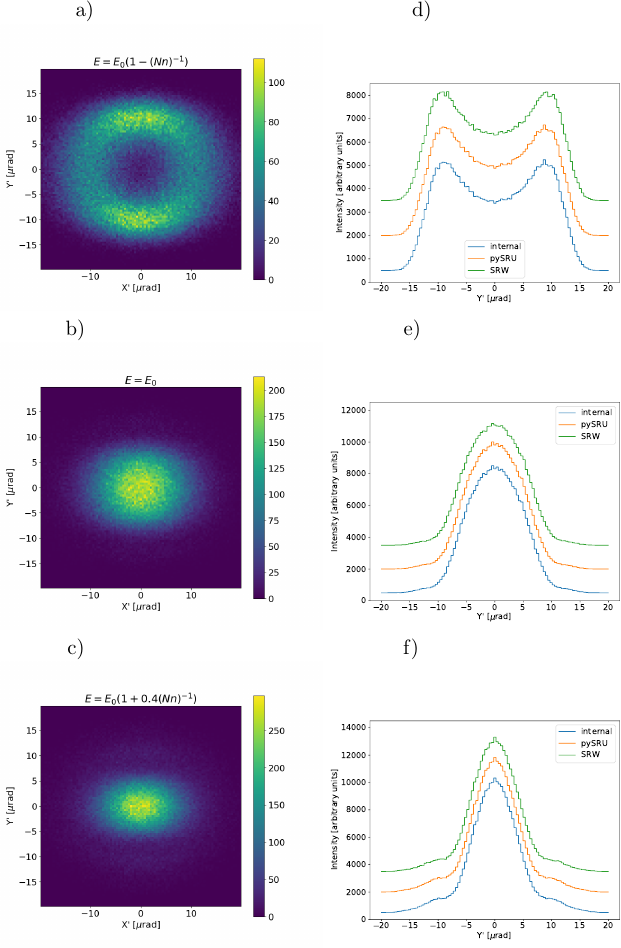}

  \caption{Divergences for ID06 U18 undulator, set close to the first harmonic resonance. a) 2D plot for photon energy red-shifted $E=E_0 [1-(Nn)^{-1}]=$\SI{9910}{eV}, b) 2D plot at resonance $E=E_0=$\SI{10000}{eV}, c) 2D plot at blue-shifted $E=E_0[1+0.4(Nn)^{-1}]=$\SI{10036}{eV} all calculated with the ``internal" algorithm. On the right, we compare the histograms of the vertical divergences given by the three calculation modes SHADOW4 internal algorithm, pySRU + WOFRY, and SRW for the three photon energies: d) red-shifted from resonance, e) at resonance, and f) blue-shifted from resonance.}
  \label{fig:red_and_blue_shifted}
\end{figure}

Another interesting feature of SHADOW4 undulators is the possibility of getting the source photon flux and use it for simulations.
In Fig.~\ref{fig:comparison_tuning_curves} we compare the flux calculated with the ``Undulator Gaussian" [that use  equation~(\ref{eq:flux_central_cone})], with the the ones obtained from OASYS add-on XOPPY that uses SRW.

Using the full undulator application ``Undulator Light Source", we first calculate the intensity distribution as a function of the vertical angle at three different photon energies (exactly on resonance, slightly red-shifted and a bit blue-shifted).
The results are in Fig.~\ref{fig:red_and_blue_shifted}, showing that the thee methods implemented in SHADOW4 produce similar results (the default ``internal" method, the use of WOFRY library, and the use of SRW to compute the radiation in the far field).

\begin{figure}
  \centering  \includegraphics[width=0.90\textwidth]{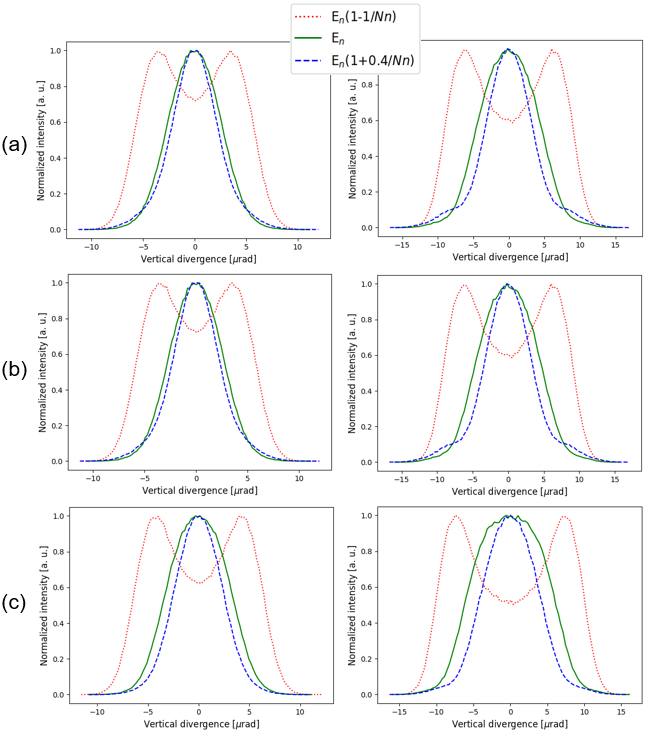}
  \caption{Left column: Vertical divergences for ID06 U18 undulator at \SI{50}{\kilo\eV} ($n$=5, in green) and two off-resonance energies (red and blue shifted), Right column: including energy spread ($\delta_\mathcal{E}$=0.001). (a) SHADOW4 internal algorithm, (b) pySRU + WOFRY, and (c) SRW. Electron beam emittace has been considered.}
  \label{fig:energy_spread__red_and_blue_shifted}
\end{figure}

In Fig.~\ref{fig:energy_spread__red_and_blue_shifted} we study the effect of the electron energy spread $\delta_\mathcal{E}$=0.001 on the vertical divergence at the $n=$ 5 harmonic. The same intensity profiles are obtained using the three modes of calculation in SHADOW4, namely the ``internal" method, the one using ``pySRU+WOFRY", and the one using SRW.

\begin{figure}
  a)~~~~~~~~~~~~~~~~~~~~~~~~~~~~~~~~~~~~~~~~~~~~~~b)~~~~~~~~~~~~~~~~~~~~~~~~~~~~~~~~~~\\\includegraphics[width=0.49\textwidth]{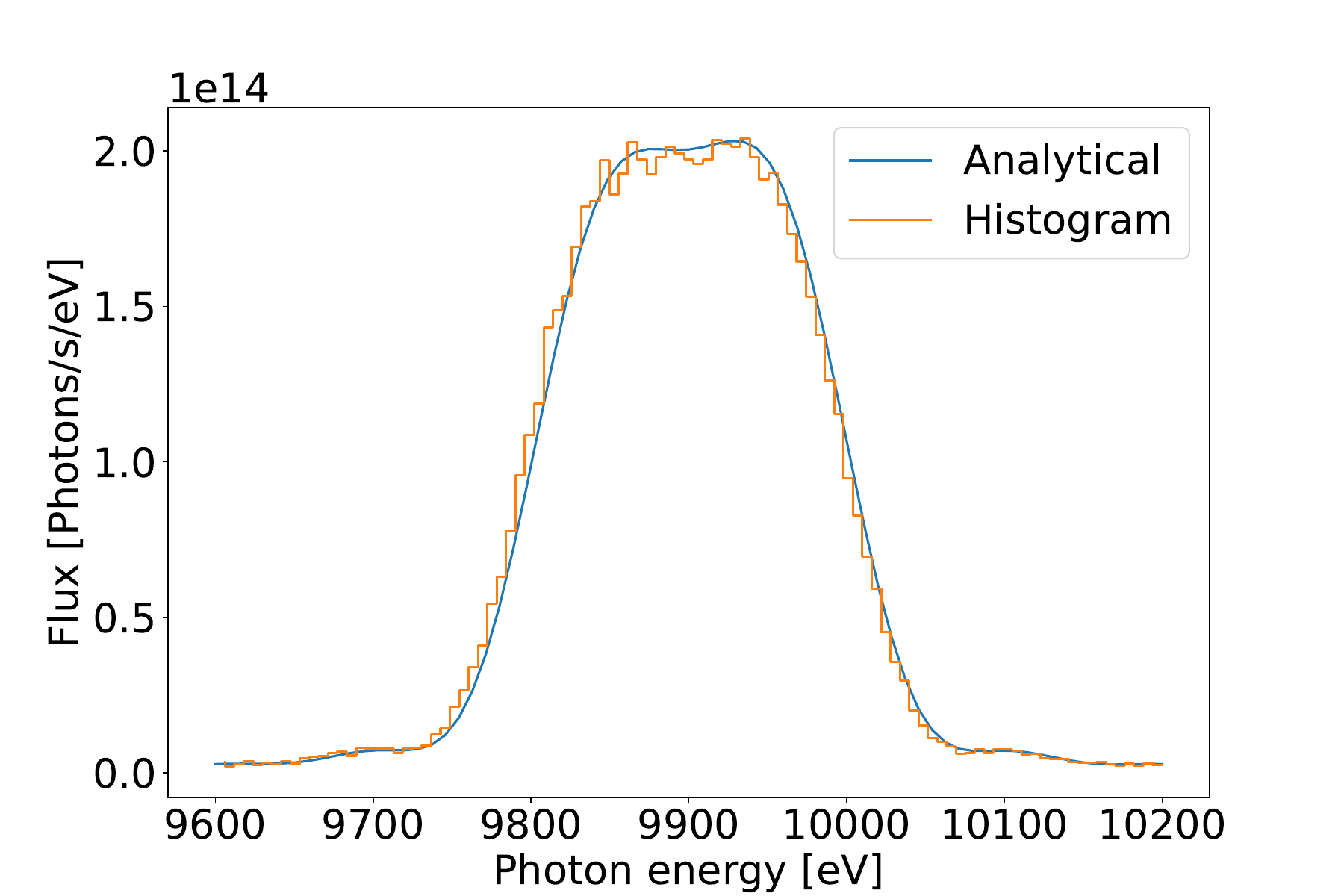}
  \includegraphics[width=0.49\textwidth]{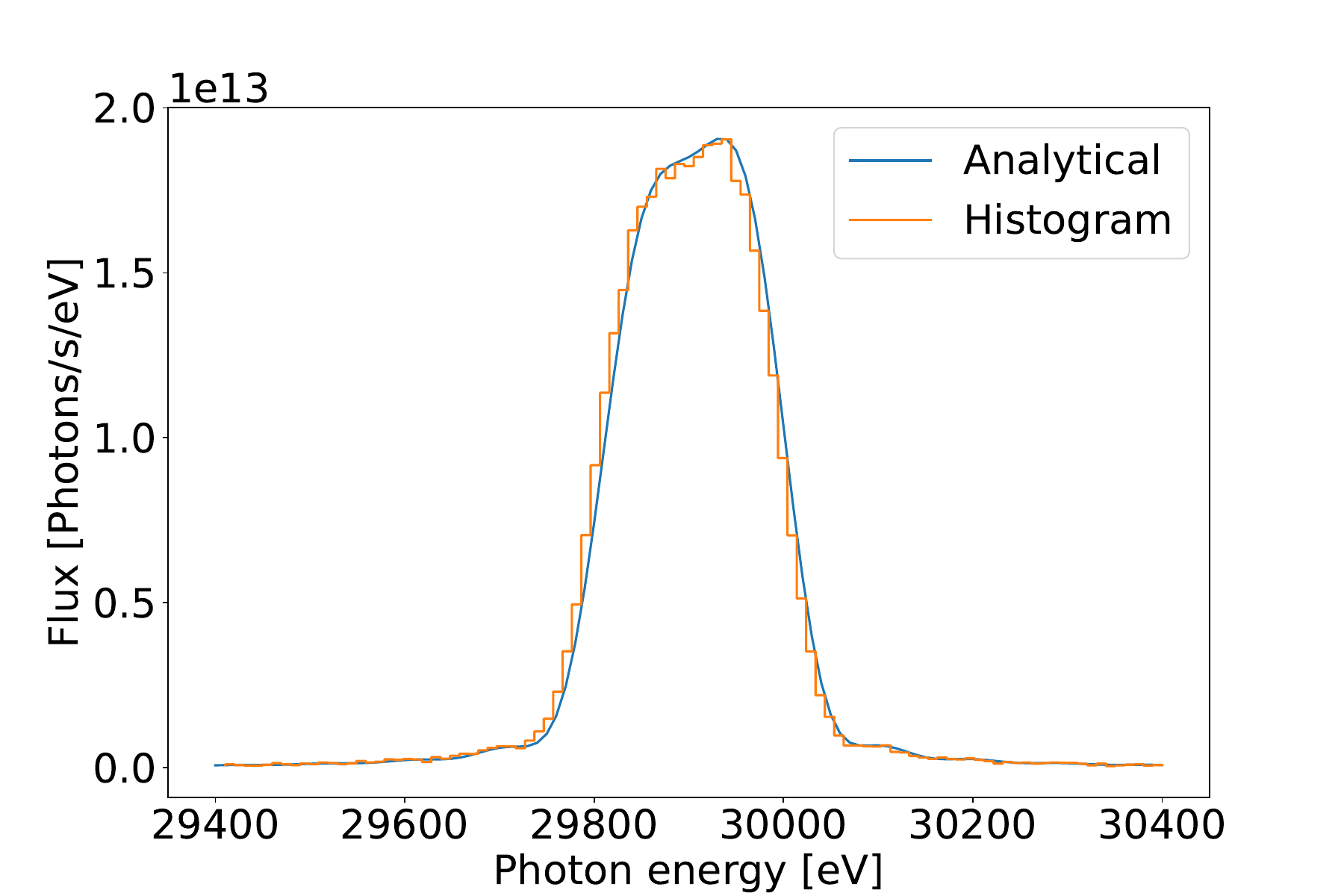}
  \caption{Energy histogram from the rays of the U18 first (a) and third harmonic (b) using undulator in poly-chromatic mode. Both are compared with the theoretical flux distribution that has been used in the sampling process.}
  \label{fig:und_flux_first_third_harm}
\end{figure}

It is often useful to simulate a polychromatic source, for example, when covering the energy range of a given undulator harmonic.
In Fig.~\ref{fig:und_flux_first_third_harm}, we display the U18 first and third harmonic histograms using the undulator in polychromatic mode (with the resonance energy at \SI{10}{\kilo\eV}). Both with 101 energy points, the first harmonic covers a range from \SI{9.6}{\kilo\eV} to \SI{10.2}{\kilo\eV} in a emission cone of \SI{16.6}{\micro\radian}, and the third harmonic from \SI{29.4}{\kilo\eV} to \SI{30.4}{\kilo\eV} in \SI{9.6}{\micro\radian}. We compared the intensity distribution from the rays with the theoretical one. 

\begin{figure}
  \centering  \includegraphics[width=0.90\textwidth]{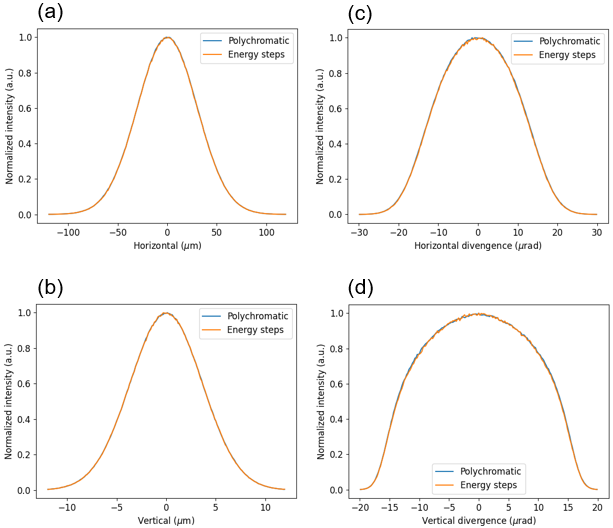}
  \caption{Comparing the poly-chromatic source with the accumulation of monochromatic steps. (a) Horizontal size, (b) vertical size, (c) horizontal divergence and (d) vertical divergence. Intensities were normalized to the intensity peak.}
  \label{fig:comparison_polychromatic_test}
\end{figure}

To verify the proper functioning of the polychromatic option, we compared the rays generated by it with those produced by multiple monochromatic sources, scanned over a range of photon energies.
This was done in 101 steps, ranging from \SI{9.6}{\kilo\eV} to \SI{10.2}{\kilo\eV}, with each energy step intensity weighted by its flux. We then compared these results with the size and divergence obtained from the polychromatic mode, same simulation from which the first harmonic in figure \ref{fig:und_flux_first_third_harm} was obtained. In figure \ref{fig:comparison_polychromatic_test}, we show the comparison results of both planes sizes and divergences profiles.

\begin{figure}
  \centering
\includegraphics[width=1.0\textwidth]{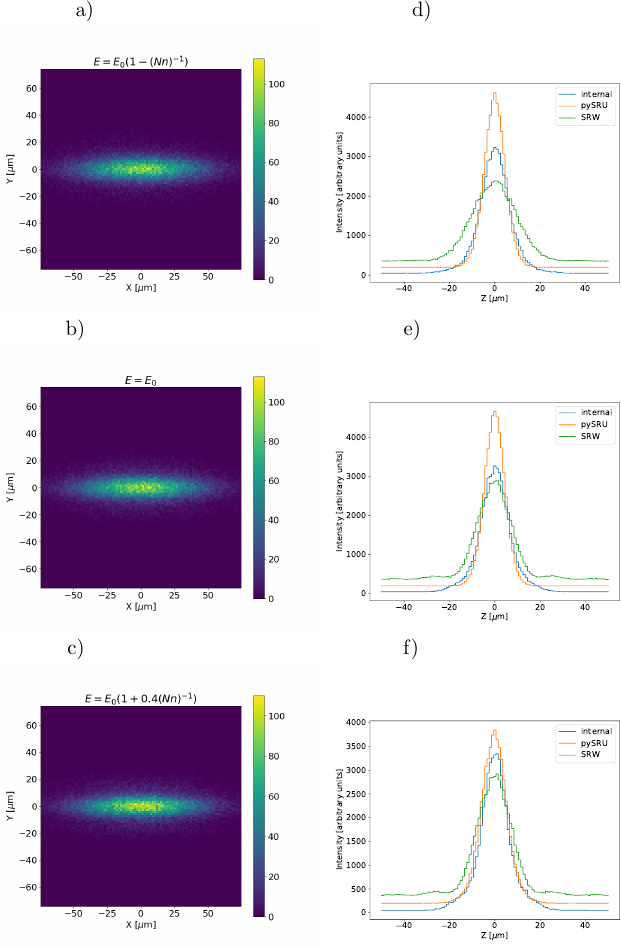}
  \caption{Sizes at the center of the ID for ID06 U18 undulator, set close to the first harmonic resonance. a) 2D plot for photon energy red-shifted $E=E_0 [1-(Nn)^{-1}]=$\SI{9910}{eV}, b) 2D plot at resonance $E=E_0=$\SI{10000}{eV}, c) 2D plot at blue-shifted $E=E_0[1+0.4(Nn)^{-1}]=$\SI{10036}{eV} all calculated with the ``internal" algorithm. On the right, we compare the histograms of the vertical size given by the three calculation modes SHADOW4 internal algorithm, pySRU + WOFRY, and SRW for the three photon energies: d) red-shifted from resonance, e) at resonance, and f) blue-shifted from resonance.}
  \label{fig:red_and_blue_backpropagated}
\end{figure}

Finally, we evaluated the diffraction limited size. Fig.~\ref{fig:red_and_blue_backpropagated}
displays the patterns at three photon energies near resonance, similar to Fig.~\ref{fig:red_and_blue_shifted}.
Slight differences are noticeable when using the various backpropagation methods implemented in SHADOW4. Fine tuning of the backpropagation parameters is necessary. We observe a significant contribution of the electron size.

\section{Summary and conclusions}
\label{sec:summary}

This work details the development and improvements of undulator sources in SHADOW4, a ray tracing code designed for synchrotron beamline modeling.
Key enhancements address critical elements for the fourth-generation synchrotron sources, including electron energy spread and diffraction-limited beam size.

A description and analysis of the existing models for the effects of the electron energy spread demonstrate the need to correct radiation divergence when using high harmonics.
This is effectively modelled by the $Q_a$ function from \cite{Tanaka2009}, as confirmed by our wave-optics numerical simulations.
The size correction is considerably less significant, primarily affecting the shape of the intensity distribution rather than its overall width.
Additionally, we found that the corrections in angle and size are not significant for photon energies outside of resonance. These insights are valuable when utilizing the new undulator features in SHADOW4: "Undulator Gaussian" and "Undulator Light Source."
 
The software tools developed here are part of the SHADOW4 add-on available in the OASYS suite \cite{codeOASYS}.
The OASYS workspaces and scripts for the simulations performed in this work are also available \cite{zenodo}


\appendix
\section{SHADOW4 methods for backpropagating the far field radiation}\label{appendix:backpropagation}

SHADOW4 stores the radiation field in a 3D stack in polar coordinates of dimension $(N_E, N_\theta, N_\varphi)$.
It is calculated at a large distance or far field (typically $D_\text{far}$=\SI{100}{\meter}).
The radiation produced by a filament electron beam, therefore it constitute a single (coherent) wavefront. 
We want to calculate the intensity map of this wavefront at the source position.
Thus we need to backpropagate the wavefront from $D$=$D_\text{far}$ to $D$=0.
Backpropagation requires the use of a propagator. 
This propagator can be implemented using the full numeric integral (which is impractical because of the long calculation time) or using using two Fourier transforms (for the Fresnel propagator).
The latter requires a cartesian gridding of the wavefront with equally distributed points.
We discarded the solution of interpolating the electric field from the polar grid $(N_\theta, N_\varphi)$ to a cartesian dense grid $(N_x,N_y)$.
The interpolation introduces high errors in the wavefront (particularly in the phase) that strongly affect the propagated wavefront.
We left for a future work the implementation of a good Fresnel propagator in polar coordinates, using the Hankel transform (instead of Fourier transform in cartesian coordinates).
Currently three solutions are available in SHADOW4: an internal solution based on backpropagation of a 1D wavefront, and other solutions based on external libraries (pySRU+WOFRY and SRW). In summary, these three methods proceed as follows:

\begin{itemize}

\item A simple solution (``internal" method) is based on reusing the original far-field $(N_E, N_\theta, N_\varphi)$ array. Without interpolation, we calculate the list of coordinates $(x,y)$ for the available points. They cover one quadrant. These points are mirrored to cover the four quadrants using the symmetry of the radiated far-field. Then the list of points and fields are propagated using a 2D integral propagator into a line on the $y$ axis.
We cannot use Fourier transforms as the points are not over a cartesian homogeneous grid. Limiting the points in the backpropagated plane to a line makes reasonable calculation times. Then, the intensity maps are calculated for each energy and then added.  Supposing axial symmetry, we can then create the 2D intensity map from the calculated 1D intensity distribution. It is from this 2D distribution map that the ray's coordinates $(x,y)$ are sampled. 

\item A second method uses the external packages pySRU and WOFRY. The far-field is re-calculated in a cartesian grid using pySRU and the 2D wavefront is backpropagated using two Fourier transforms with WOFRY.
The obtained backpropagated fields (one for each photon energy) are used to compute the intensities.
The intensity map for the whole energy interval (obtained by adding the intensities of the individual photon energies) is used to sample the $(x,y)$ coordinates of the rays. 

\item A third solution, similar to the previous one, uses SRW \cite{codeSRW} for calculating the source and backpropagating it. All calculations are done in cartesian coordinates.

\end{itemize}



\ack{\textbf{Acknowledgements}}

We acknowledge Luca Rebuffi and Xianbo Shi (Argonne National Laboratory) for helpful discussions and alpha-test the code. We are in debt with Boaz Nash for fruitfull discussions and giving us access to the data in his paper \cite{Nash2019}. We would like to acknowledge as well Takashi Tanaka for his help on the use of SPECTRA software and sharing his insights on the energy spread effect over in both source size and divergence.

\newpage
\referencelist{iucr}



@misc{zenodo,
  author       = {Manuel {Sanchez del Rio} and
                  Reyes-Herrera, Juan},
  title        = {{oasys-esrf-kit/paper-undulators-resources: Scripts 
                   at the submission}},
  month        = oct,
  year         = 2024,
  publisher    = {Zenodo},
  version      = {v1.1},
  doi          = {https://doi.org/10.5281/zenodo.13940373},
  url          = {https://doi.org/10.5281/zenodo.13940373},
  howpublished = "\url{https://doi.org/10.5281/zenodo.13940373}",
}

@book{elleaume,
  title={Undulators, Wigglers and Their Applications},
  author={Onuki, H. and Elleaume, P.},
  isbn={9780203218235},
  year={2003},
  publisher={CRC Press}
}

@inbook{ elleaumeChapter3,
author = {P. Elleaume},
chapter = {3: Undulator Radiation},
crossref = {elleaume}
}

@inbook{Eckhardt,
author = {R. Eckhardt},
chapter = {Stan Ulam, John von Neumann, and the Monte Carlo Method},
crossref = {book_mcnp},
url={https://permalink.lanl.gov/object/tr?what=info:lanl-repo/lareport/LA-UR-88-9068}
}

@book{book_mcnp,
  title={Los Alamos Science, special issue, Stanislaw Ulam 1909-1984},
  author={},
  isbn={},
  year={1987},
  publisher={Los Alamos National Laboratiry},
  url={https://la-science.lanl.gov/lascience15.shtml}
}

@book{jackson,
  title={Classical Electrodynamics},
  author={Jackson, J. D.},
  isbn={0-471-30932-X},
  year={1999},
  publisher={Wiley}
}

@article{kim1986b,
title = "Brightness, coherence and propagation characteristics of synchrotron radiation",
journal = "Nuclear Instruments and Methods in Physics Research Section A: Accelerators, Spectrometers, Detectors and Associated Equipment",
volume = "246",
number = "1",
pages = "71 - 76",
year = "1986",
issn = "0168-9002",
doi = "10.1016/0168-9002(86)90048-3",
url = "https://doi.org/10.1016/0168-9002(86)90048-3",
author = "Kwang-Je Kim"
}

@misc{pySRU,
  author = {Thery, S. and Glass, M. and and Sanchez del Rio, M},
  title = {{PySRU} {G}it{H}ub repository in \\ \textit{https://www.github.com/oasys-kit/pySRU}},
  year = {2016},
  publisher = {GitHub},
  journal = {GitHub repository},
  url  = 	 {https://github.com/oasys-kit/pySRU}
}

@article{codeSRW,
    author = {O. Chubar and P. Elleaume},
    title = {Accurate and efficient computation of synchrotron radiation in the near field region},
    journal = {Proceedings of the 6th European Particle Accelerator Conference - EPAC-98},
    pages = {1177--1179},
    url = {https://accelconf.web.cern.ch/e98/papers/THP01G.pdf}, year = {1998},
}

@article{codeSHADOW,
    author = {Sanchez del Rio, M. and Canestrari, N. and Jiang, F. and Cerrina, F.},
    title = {{\it SHADOW3}: a new version of the synchrotron X-ray optics modelling package},
    journal = {Journal of Synchrotron Radiation},
    url = {http://dx.doi.org/10.1117/12.893433},
    year = {2011},
    volume = {18},
    number = {5},
    pages = {708--716}
}

@article{codeSHADOWOUI,
    author = {Rebuffi, L. and Sanchez del Rio, M.},
    title = {{\it ShadowOui}: a new visual environment for X-ray optics and synchrotron beamline simulations},
    journal = {Journal of Synchrotron Radiation},
    url={https://doi.org/10.1107/S1600577516013837},
    year = {2016},
    volume = {23},
    number = {6},
    pages = {1357--1367},
    month = "November"
}

@article{codeOASYS,
  author = {Rebuffi, L. and Sanchez del Rio, M.},
  title = {{OASYS (OrAnge SYnchrotron Suite): an open-source graphical environment for x-ray virtual experiments}},
  journal = {Proc. SPIE 10388: Advances in Computational Methods for X-Ray Optics IV},
  volume = {10388},
  pages = {169--177},
  year = {2017},
  doi = {10.1117/12.2274263},
  URL = {http://doi.org/10.1117/12.2274263}
}

@article{codeWOFRY,
    author = {Rebuffi, Luca and {Sanchez del Rio}, Manuel},
    title = {{Interoperability and complementarity of simulation tools for beamline design in the OASYS environment}},
    volume = {10388},
    journal = {Proc. SPIE 10388, Advances in Computational Methods for X-Ray Optics IV},
    organization = {International Society for Optics and Photonics},
    pages = {28 -- 41},
    year = {2017},
    doi = {10.1117/12.2274232},
    URL = {https://doi.org/10.1117/12.2274232}
}

@article{codeURGENT,
  doi = {10.1063/1.1142766},
  url = {https://doi.org/10.1063/1.1142766},
  year = {1992},
  month = jan,
  publisher = {{AIP} Publishing},
  volume = {63},
  number = {1},
  pages = {392--395},
  author = {R. P. Walker and B. Diviacco},
  title = {{URGENT}-A computer program for calculating undulator radiation spectral,  angular,  polarization,  and power density properties},
  journal = {Review of Scientific Instruments}
}

@article{codeUS,
  doi = {10.1016/0168-9002(94)91855-4},
  url = {https://doi.org/10.1016/0168-9002(94)91855-4},
  year = {1994},
  month = aug,
  publisher = {Elsevier {BV}},
  volume = {347},
  number = {1-3},
  pages = {61--66},
  author = {Roger J. Dejus and Alfredo Luccio},
  title = {Program {UR}: General purpose code for synchrotron radiation calculations},
  journal = {Nuclear Instruments and Methods in Physics Research Section A: Accelerators,  Spectrometers,  Detectors and Associated Equipment}
}

@article{Tanaka2001,
    author = "Tanaka, Takashi and Kitamura, Hideo",
    title = "{{\it SPECTRA}: a synchrotron radiation calculation code}",
    journal = "Journal of Synchrotron Radiation",
    year = "2001",
    volume = "8",
    number = "6",
    pages = "1221--1228",
    doi = {10.1107/S090904950101425X},
    url = {https://doi.org/10.1107/S090904950101425X},
}

@article{geloni2008,
title = "Transverse coherence properties of X-ray beams in third-generation synchrotron radiation sources",
journal = "Nucl. Instr. and Meth. A",
volume = "588",
number = "3",
pages = "463 - 493",
year = "2008",
note = "",
issn = "0168-9002",
doi = "j.nima.2008.01.089",
url = "http://doi.org/10.1016/j.nima.2008.01.089",
author = "Gianluca Geloni and Evgeni Saldin and Evgeni Schneidmiller and Mikhail Yurkov"
}

@article{Geloni2018,
  title = {Effects of energy spread on brightness and coherence of undulator sources},
  volume = {25},
  ISSN = {1600-5775},
  url = {http://dx.doi.org/10.1107/S1600577518010330},
  DOI = {10.1107/s1600577518010330},
  number = {5},
  journal = {Journal of Synchrotron Radiation},
  publisher = {International Union of Crystallography (IUCr)},
  author = {Geloni,  Gianluca and Serkez,  Svitozar and Khubbutdinov,  Ruslan and Kocharyan,  Vitali and Saldin,  Evgeni},
  year = {2018},
  month = aug,
  pages = {1335–1345}
}

@article{Walker2019,
  title = {Undulator radiation brightness and coherence near the diffraction limit},
  author = {Walker, Richard P.},
  journal = {Phys. Rev. Accel. Beams},
  volume = {22},
  issue = {5},
  pages = {050704},
  numpages = {12},
  year = {2019},
  month = {May},
  publisher = {American Physical Society},
  doi = {10.1103/PhysRevAccelBeams.22.050704},
  url = {https://link.aps.org/doi/10.1103/PhysRevAccelBeams.22.050704}
}

@article{Tanaka2014,
  title = {Numerical methods for characterization of synchrotron radiation based on the Wigner function method},
  author = {Tanaka, Takashi},
  journal = {Phys. Rev. ST Accel. Beams},
  volume = {17},
  issue = {6},
  pages = {060702},
  numpages = {14},
  year = {2014},
  month = {Jun},
  publisher = {American Physical Society},
  doi = {10.1103/PhysRevSTAB.17.060702},
  url = {https://link.aps.org/doi/10.1103/PhysRevSTAB.17.060702}
}

@article{krinsky,
  title = {Undulators as a source of synchrotron radiation},
  author = {Krinsky, Samuel},
  journal = {IEEE Transactions on Nuclear Science},
  volume = {NS-30},
  issue = {4},
  pages = {3078-3082},
  numpages = {5},
  year = {1983},
  month = {August}
}

@article{kim1986a,
title = "Angular distribution of undulator power for an arbitrary deflection parameter K",
journal = "Nuclear Instruments and Methods in Physics Research Section A: Accelerators, Spectrometers, Detectors and Associated Equipment",
volume = "246",
number = "1",
pages = "67 - 70",
year = "1986",
issn = "0168-9002",
doi = "10.1016/0168-9002(86)90047-1",
url = "https://doi.org/10.1016/0168-9002(86)90047-1",
author = "Kwang-Je Kim"
}

@inbook{ KimInXraydatabooklet,
author = {K-J Kim},
chapter = {Characteristics of Synchrotron Radiation},
crossref = {xraydatabooklet}
}

@book{xraydatabooklet,
  title={X-ray Data Booklet},
  author={Thompson, A.C.},
  url={http://xdb.lbl.gov/},
  year={2001},
  publisher={Lawrence Berkeley National Laboratory, University of California}
}

@article{kim1989, 
  title={Characteristics of synchrotron radiation}, 
  ISSN={0094-243X}, 
  url={http://doi.org/10.1063/1.38046}, 
  DOI={10.1063/1.38046}, 
  journal={AIP Conference Proceedings}, 
  publisher={AIP}, 
  author={Kim, Kwang-Je},
  year={1989}}

@article{lindberg2015,
  title = {Compact representations of partially coherent undulator radiation suitable for wave propagation},
  author = {Lindberg, Ryan R. and Kim, Kwang-Je},
  journal = {Phys. Rev. ST Accel. Beams},
  volume = {18},
  issue = {9},
  pages = {090702},
  numpages = {16},
  year = {2015},
  month = {Sep},
  publisher = {American Physical Society},
  url = {http://doi.org/10.1103/PhysRevSTAB.18.090702},
  doi = {10.1103/PhysRevSTAB.18.090702}
}

@article{borland2012,
  title={ Progress towards ultimate storage ring light sources },
  url={https://accelconf.web.cern.ch/accelconf/IPAC2012/papers/tuxb01.pdf},
  journal={Proceedings of IPAC2012, New Orleans, Louisiana, USA},
  publisher={ IPAC'12 OC / IEEE},
  author={Borland, M.},
  year={2012},
  month={July}}

@article{hettel2014,
  title={ Chellenges in design of diffraction limited storage rings },
  url={http://accelconf.web.cern.ch/AccelConf/IPAC2014/papers/moxba01.pdf},
  journal={IPAC2014: Proceedings of the 5th International Particle Accelerator Conference},
  publisher={ IPAC'14},
  author={Hettel, R.},
  year={2014},
  month={July}}

@article{huang2013,
  title={BRIGHTNESS AND COHERENCE OF SYNCHROTRON RADIATION AND FELs∗},
  url={https://accelconf.web.cern.ch/accelconf/IPAC2013/papers/moycb101.pdf},
  journal={IPAC2013: Proceedings of the 4th International Particle Accelerator Conference},
  publisher={ IPAC'13},
  author={Huang, Z.},
  year={2013},
  month={June}}

@article{shadow2undulators,
  title={Modelling of undulator sources},
  author={Chapman, K and Lai, B and Cerrina, F and Viccaro, J},
  journal={Nuclear Instruments and Methods in Physics Research Section A: Accelerators, Spectrometers, Detectors and Associated Equipment},
  volume={283},
  number={1},
  pages={88--99},
  year={1989},
  publisher={Elsevier}
}

@article{Tanaka2009,
author = "Tanaka, Takashi and Kitamura, Hideo",
title = "{Universal function for the brilliance of undulator radiation considering the energy spread effect}",
journal = "Journal of Synchrotron Radiation",
year = "2009",
volume = "16",
number = "3",
pages = "380--386",
month = "May",
doi = {10.1107/S0909049509009479},
url = {https://doi.org/10.1107/S0909049509009479},
}

@misc{ wiki_erf,
    author = "{Wikipedia contributors}",
    title = "Error function --- {Wikipedia}{,} The Free Encyclopedia",
    year = "2024",
    url = "https://en.wikipedia.org/w/index.php?title=Error\_function\&oldid=1245988545",
    howpublished = "\url{https://en.wikipedia.org/w/index.php?title=Error\_function\&oldid=1245988545}",
    note = "[Online; accessed 9-October-2024]"
  }

@inproceedings{Nash2019,
author = {B. Nash and O. Chubar and D. L. Bruhwiler and M. Rakitin and P. Moeller and R. Nagler and N. Goldring},
title = {{Undulator radiation brightness calculations in the Sirepo GUI for SRW}},
volume = {11110},
booktitle = {Advances in Laboratory-based X-Ray Sources, Optics, and Applications VII},
editor = {Alex Murokh and Daniele Spiga},
organization = {International Society for Optics and Photonics},
publisher = {SPIE},
pages = {111100K},
year = {2019},
doi = {10.1117/12.2530663},
URL = {https://doi.org/10.1117/12.2530663}
}

@inproceedings{Cerrina1984,
  doi = {10.1117/12.944815},
  url = {https://doi.org/10.1117/12.944815},
  year = {1984},
  month = dec,
  publisher = {{SPIE}},
  author = {F. Cerrina},
  editor = {Jeremy M. Lerner},
  title = {Ray Tracing Of Recent {VUV} Monochromator Designs},
  booktitle = {{SPIE} Proceedings}
}

@article{ShadowSRN2023,
author = {Sanchez del Rio, M. and Rebuffi, L.},
title = {40 Years of SHADOW: Serving Four Generations of Synchrotron Facilities},
journal = {Synchrotron Radiation News},
volume = {36},
number = {5},
pages = {6-7},
year = {2023},
publisher = {Taylor & Francis},
doi = {10.1080/08940886.2023.2274745},
URL = { 
        https://doi.org/10.1080/08940886.2023.2274745
},
eprint = { 
        https://doi.org/10.1080/08940886.2023.2274745
}
}

@book{BookDuke,
	author = {Duke, Philip John},
	title = {Synchrotron radiation :},
	publisher = {Oxford Univ. Press,},
	year = {2000},
	series = {Oxford series on synchrotron radiation ;},
	address = {Oxford [u.a.] :},
	edition = {1. publ.},
	note = {Includes index. - Bibliography},
	url = {http://www.loc.gov/catdir/enhancements/fy0611/00703272-t.html}
}

@book{BookClarke,
	author = {Clarke James A.},
	title = {The science and technology of undulators and wigglers /},
	publisher = {Oxford University Press,},
	year = {2004},
	series = {Oxford series on synchrotron radiation ;},
	address = {Oxford :},
	note = {Includes bibliographical references and index. - Description based on print version record},
	url = {http://dx.doi.org/10.1093/acprof:oso/9780198508557.001.0001}
}
\end{document}